\documentclass[aps,pra,twocolumn,footinbib,notitlepage,floatfix]{revtex4-2}
\usepackage{amsmath}
\usepackage{amssymb}
\usepackage{graphicx}
\usepackage{dcolumn}
\usepackage{enumerate}
\usepackage{bm}
\usepackage[normalem]{ulem}
\usepackage{xcolor}
\usepackage{comment}
\usepackage{subfigure}
\usepackage{breqn}
\usepackage{hyperref}
\usepackage{braket}
\usepackage{orcidlink}
\usepackage[sectionbib]{bibunits}
\defaultbibliographystyle{apsrev4-2}
\defaultbibliography{library}
\hypersetup{colorlinks=true, citecolor=blue, urlcolor=blue, linkcolor=blue}
\usepackage{environ}
\NewEnviron{equationS}{%
\begin{align}\begin{split}
  \BODY
\end{split}\end{align}
}

\newcommand{\dd}{{\rm d}}
\newcommand{\ghost}[1]{{\color{red}{}\normalcolor}}

\makeatletter
\let\cat@comma@active\@empty
\newcommand*{\balancecolsandclearpage}{%
  \close@column@grid
  \clearpage
  \twocolumngrid
}
\makeatother

\begin{document}
\title{A ``squeezed polaron" variational wavefunction for the spin boson model
}
\author{Diego Barberena~\orcidlink{0000-0002-6845-1807}}
\email{db985@cam.ac.uk}
\affiliation{T.C.M. Group, Cavendish Laboratory, University of Cambridge, J.J. Thomson Avenue, Cambridge CB3 0US, UK}
\author{Nigel Cooper~\orcidlink{0000-0002-4662-1254}}
\affiliation{T.C.M. Group, Cavendish Laboratory, University of Cambridge, J.J. Thomson Avenue, Cambridge CB3 0US, UK}
\date{\today}

\begin{abstract}
The localization transition of the sub-Ohmic spin-boson model generates boson bath correlations that are, at low frequencies, analytically inaccessible to standard coherent-state-polaron variational ansatzes. In this paper, we introduce a ``squeezed polaron'' wavefunction incorporating generic Gaussian boson-boson correlations induced by the spin impurity. This gives the correct critical power-law scaling of bath observables near the localization transition together with a systematically improved ground state energy and a more accurate determination of the critical coupling. Our wavefunction captures the correct mean-field nature of the transition in the deep sub-Ohmic region. It also displays non-mean-field critical exponents in the shallow sub-Ohmic regime. Using the squeezed polaron as a starting point, we derive scattering phase shifts for bosons, and show how they encode the emergent energy scale that vanishes at the localization transition.
\end{abstract}

\maketitle
{\it Introduction:} The spin-boson model~\cite{Leggett1987,Weiss2012} is an iconic model of quantum dissipation that describes the interaction between a single two-level system (TLS) and an environmental bath of bosonic modes. It thus provides a minimal setting in which to study phenomena such as decoherence and relaxation across a wide range of parameter regimes, including those characterized by the presence of bath-induced memory effects~\cite{Clos2012,backer2025verifyingquantummemorydynamics} and strong system-environment coupling~\cite{Peropadre2013,Forn-Diaz2017}. The model is also of experimental relevance, both as an effective description of e.g. dissipative quantum tunneling in a variety of systems~\cite{Golding1992,Chakravarty1984,Stockburger1994,Han1991}, and as the target of quantum simulation efforts~\cite{Magazzu2018,Sun2025}.

Despite the simplicity of its ingredients, the spin-boson model displays a rich phenomenology. In its ground state, it can exhibit a second-order quantum phase transition~\cite{Bulla2003,Bulla2005}. This transition is accessed by increasing the coupling between the TLS and its environment, which is parametrized by a spectral function $J(\omega)\propto \alpha\,\omega^s$. In addition to an overall interaction strength $\alpha$, $J(\omega)$ depends on the exponent $s$, which controls how strongly the lowest-frequency modes interact with the TLS. When $s\leq1$ (sub-Ohmic regime) and $\alpha$ is small, the ground state is unique and the TLS has zero magnetization, giving rise to the ``delocalized" phase. At larger $\alpha$, the system enters a ``localized" phase with two degenerate ground states in which the TLS has opposite magnetizations and the low frequency bosons have infinite occupation. Owing to the simplicity of the model, this transition has been extensively studied using a diverse set of numerical techniques~\cite{Winter2009,2012Guo,Florens2011,Blunden-Codd2017,SHEN20235}.

The model has also been tackled using analytics~\cite{Florens2011,Florens2010}. In particular, the variational principle has been very successful in capturing the key physics of the spin-boson ground state. Polaron ansatzes~\cite{SilbeyHarris1984,Chin2011,Zhao2011}, whereby the TLS's spin up and down states are each dressed by a finite number of superposed bosonic coherent states (the ``polarons"), incorporate efficiently spin-bath correlations. Furthermore, they serve as the starting point of refinements designed to capture other expected features of the ground state~\cite{Nazir2012,Zheng2013,He2018,Bera2014b}, are at the basis of more exhaustive numerical approaches~\cite{Bera2014}, and are also useful to study dynamical processes such as scattering~\cite{Bera2016,DiazCamacho2016,Shi2018}, direct driving~\cite{McCutcheon_2010}, quantum state preparation~\cite{Bond2024}, etc. At the same time, a finite number of coherent-state polarons cannot~\cite{Blunden-Codd2017} capture accurately the boson-boson correlations that develop among the lowest-frequency bosons in the vicinity of the localization transition.

In this paper, we replace the coherent state polaron ansatz by an ansatz that allows generic Gaussian boson-boson correlations. Previously such correlations have only been considered numerically in the Ohmic limit  ($s=1$)~\cite{SHI2018b}. We find two solution branches, corresponding to the localized and delocalized phases. We show analytically that the delocalized branch displays the correct low-energy scaling laws for boson observables~\cite{Blunden-Codd2017}. Furthermore, the values of the critical coupling derived from our wavefunction are close to those obtained via numerically exact approaches, and coincide, when $s\lesssim 0.4$, with the values of the coupling at which excitations of the bosonic cloud surrounding the spin becomes gapless. The model also provides better accuracy for physical observables, especially in the shallow sub-Ohmic regime ($0.5<s<1$). Intriguingly, in the delocalized branch our ``squeezed" polaron also reflects the change from mean-field to non mean-field behaviour as $s$ moves from the deep ($0<s<0.5$) to the shallow sub-Ohmic regions, although the critical exponents we find in the latter are not numerically accurate and they occur in a parameter region where the localized branch is at a lower energy. To conclude the paper we predict the scattering properties of bosons from the squeezed polaron background. By analyzing these features across a broad frequency range, we show how environmental probes can be used to extract deeper information about the system.

\begin{figure}
    \centering
    \includegraphics[width=0.98\linewidth]{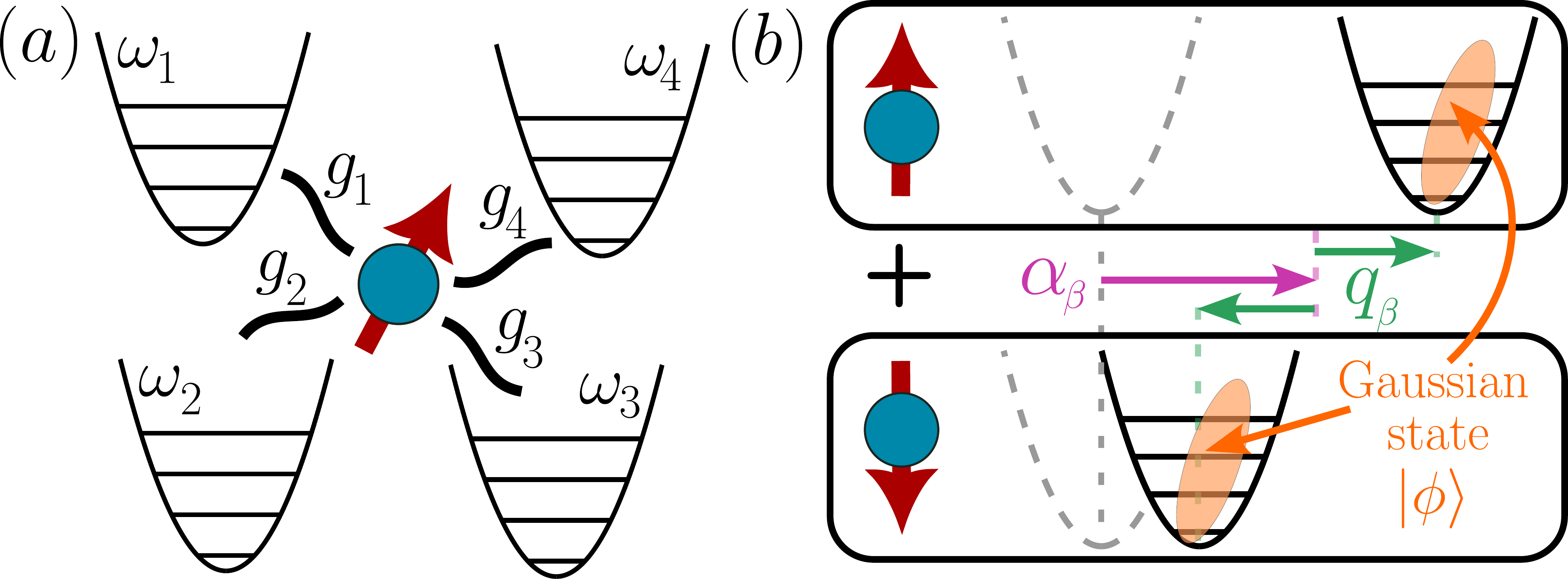}
    \caption{(a) The spin boson model describes a single two-level system interacting with an environment of bosonic modes. (b) Graphical depiction of our ansatz wavefunction as a superposition which involves correlated ($q_\beta$) and uncorrelated ($\alpha_\beta$) displacements of bosonic modes in a gaussian state $\ket{\phi}$ and a rotation of the two-level system (not shown, manifests in the relative weights of the two states).}
    \label{fig:Schematic}
\end{figure}

\noindent\textit{Model:}
The spin-boson model is defined by the following Hamiltonian:
\begin{equation}\label{eqn:Model:Hamiltonian}
    \hat{H}=\Delta\hat{s}_x+\sum_\beta g_\beta(\hat{a}_\beta+\hat{a}_\beta^\dagger)\hat{s}_z+\sum_\beta\omega_\beta\hat{a}_\beta^\dagger\hat{a}_\beta,
\end{equation}
where $\hat{s}_{x,y,z}=\hat{\sigma}_{x,y,z}/2$ and $\hat{\sigma}_{x,y,z}$ are the standard Pauli operators acting on the two-level system (TLS), $\hat{a}_\beta/\hat{a}_\beta^\dagger$ are bosonic annihilation/creation operators for bath mode $\beta$ with frequency $\omega_\beta$, $\Delta$ is the tunneling amplitude of the TLS, $g_\beta$ is the coupling between the TLS and bath mode $\beta$ [see Fig.~\ref{fig:Schematic}(a)] and we set $\hbar=1$. We also introduce $\hat{p}_\beta=-i(\hat{a}_\beta-\hat{a}_\beta^\dagger)/\sqrt{2}$ and $\hat{x}_\beta=(\hat{a}_\beta+\hat{a}_\beta^\dagger)/\sqrt{2}$.

The effects of the bath on the TLS are determined by the spectral function $J(\omega)=\pi\sum_\beta g_\beta^2 \delta(\omega-\omega_\beta)$~\cite{Leggett1987}. For our sub-Ohmic bath $J(\omega)=2\pi\alpha\omega_c^{1-s} \omega^s \theta(\omega_c-\omega)$ is specified by a cutoff frequency $\omega_c\gg\Delta$, a coupling strength $\alpha$ and the bath exponent $s$. When $s\leq 1$, the system undergoes a localization transition at a critical value of $\alpha$, from a delocalized phase at small $\alpha$ with $\braket{\hat{s}_z}=0$, to a localized phase at larger $\alpha$ with $\braket{\hat{s}_z}\neq 0$. The transition is in the mean field universality class when $0<s\leq0.5$ but is non mean field when $0.5<s<1$.

A large class of variational wavefunctions for Eq.~(\ref{eqn:Model:Hamiltonian}) can be written down by expressing the ground state $\ket{\psi}$ as $\ket{\psi}=\hat{U}\ket{\phi}$, where $\ket{\phi}$ is a simple state and 
\begin{equation}
    \hat{U}=\hat{\mathcal{D}}_c\hat{\mathcal{D}}_u\hat{R}(\theta)
\end{equation}
is a unitary composed of three operations. The first operation $\hat{R}(\theta)=\exp(-i\theta\hat{s}_y)$ rotates the spin about the $y$ direction. The second operation $\hat{\mathcal{D}}_u=\exp\big[-i\sqrt{2}\sum_\beta\alpha_\beta\hat{p}_\beta\big]$ displaces the bath modes along the $x_\beta$ direction in phase space; the final operation $\hat{\mathcal{D}}_c=\exp\big[i\sqrt{2}\hat{s}_z\sum_\beta q_\beta\hat{p}_\beta\big]$ also displaces the bath modes along $x_\beta$, but now in a direction correlated with the $z$ projection of the spin state [see Fig.~\ref{fig:Schematic}(b)]. Physically, $\hat{\mathcal{D}}_c$ introduces spin-bath entanglement and correlations by dressing the up and down spin components with different bosonic clouds.

Within the framework defined by $\hat{U}$, the celebrated Silbey-Harris~\cite{SilbeyHarris1984} wavefunction results from setting $\alpha_\beta=\theta=0$, choosing $\ket{\phi_0}=\ket{-x}\otimes\ket{0} $ to be a product of a spin polarized along $-x$ and the boson vacuum, and minimizing over $q_\beta$. The generalization developed by Chin~\cite{Chin2011} uses the same $\ket{\phi_0}$, but minimizes over $\alpha_\beta,\theta, q_\beta$ in order to capture both phases of the system.

\noindent\textit{The squeezed polaron:} Our refinement consists in allowing $\ket{\phi}$ to be a generic gaussian state of the bosonic bath, meaning that it is entirely characterized by the two-point correlators of $\hat{x}_\beta,\hat{p}_\beta$ via Wick's theorem. The minimization procedure will then select the optimal bath-bath correlations induced by the presence of the TLS. Using the gaussianity properties of $\ket{\phi}$ we obtain the following ground state energy
\begin{align}\begin{split}
    E&=-\frac{\cos\theta}{2}\overbrace{\Delta e^{-\sum_{\beta\beta'}q_\beta q_{\beta'}\braket{\hat{p}_\beta\hat{p}_{\beta'}}_\phi}}^{D}+\sum_{\beta}\omega_\beta\braket{\hat{a}_\beta^\dagger\hat{a}_\beta}_\phi\\
    &+\sum_\beta \omega_\beta\alpha_\beta^2+\sum_\beta (g_\beta-\omega_\beta q_\beta)\alpha_\beta \sin\theta\\
    &+\sum_\beta\frac{(g_\beta-\omega_\beta q_\beta)^2}{4\omega_\beta}-\sum_\beta\frac{g_\beta^2}{4\omega_\beta},
\end{split}\end{align}
where the average $\braket{}_\phi$ is taken with respect to the gaussian state $\ket{\phi}$ and we have introduced a renormalized coherence $D$ as indicated by the overbrace. Note also that $E$ depends only on two-point correlators of $\ket{\phi}$. The general minimization procedure is detailed in the Supplementary Material~\cite{SM}. The end result is that, at a fixed angle $\theta$, the displacements become functions of the scaled mode frequencies $y_\beta=\omega_\beta\cos\theta/D$ and are given by 
\begin{align}\label{eqn:SqueezedPolaron:Displacements}
    \begin{split}
        q_\beta&=\frac{g_\beta\cos\theta/D}{y_\beta+m(y_\beta)},\hspace{1cm}\alpha_\beta=-\frac{q_\beta \,m(y_\beta)\sin\theta }{2y_\beta}
    \end{split}\,,
\end{align}
where $m(y)$ is a function satisfying the integral equation
\begin{equation}\label{eqn:SqueezedPolaron:IntegralM(y)}
    m(y)=\int_0^{\infty}\frac{2y\, \dd z/\pi}{y^2+z^2}\frac{1}{1+2b\int_0^{y_c}\frac{ x^{s+1}\,\dd x}{(x^2+z^2)\big[x+m(x)\big]^2}},
\end{equation}
with $y_c=\omega_c\cos\theta/D$. The parameter $b$ in Eq.~(\ref{eqn:SqueezedPolaron:IntegralM(y)}) is a function of the scaled coupling constant $\alpha_\Delta=\alpha(\omega_c/\Delta)^{1-s}$, defined implicitly via the relation
\begin{equation}\label{eqn:SqueezedPolaron:CouplingConstant}
    \alpha_\Delta=\frac{b}{(\cos\theta)^{3-s}}\exp\left(-\frac{b(1-s)}{(\cos\theta)^2} \int_0^{y_c}\frac{m(y)y^s \,\dd y}{\big[y+m(y)\big]^2}\right),
\end{equation} 
For comparison, the Silbey-Harris and Chin results are obtained by setting $m(y)\to 1$ in Eq.~(\ref{eqn:SqueezedPolaron:Displacements}) and Eq.~(\ref{eqn:SqueezedPolaron:CouplingConstant}), meaning that our ansatz allows for more complex behaviour of the displacements at low frequency. Standard observables like the renormalized coherence $2\braket{\hat{s}_x}=D$ and ground state energy are given by
\begin{align}\begin{split}\label{eqn:SqueezedPolaron:Observables}
    D&=\Delta\exp\left(-\frac{b}{(\cos\theta)^2}\int_0^{y_c}\frac{m(y)y^s \,\dd y}{\big[y+m(y)\big]^2}\right)\\[5pt]
    \delta E&=-\frac{D\cos\theta}{2}\left[1-\frac{b(1-s)}{(\cos\theta)^2}\int_0^{y_c}\frac{y^{s-1} m(y)\,\dd y}{y+m(y)}\right].
\end{split}\end{align}
where $\delta E=E-\alpha \omega_c/(2s)$ is the energy gain with respect to the adiabatic approximation~\cite{Weiss2012}.  Minimization over $\theta$ then leads either to $\theta=0$ (delocalized branch) or
\begin{equation}\label{eqn:SqueezedPolaron:Angle}
    (\cos\theta)^2=2b\int_0^{y_c}\frac{y^{s-1}m(y)^2\,\dd y}{\big[y+m(y)\big]^2}
\end{equation}
Taken together as a whole, these equations define $D$, $\delta E$, $\theta$, and $\alpha_\Delta$  as parametric functions of $b$ but the true independent variable, constructed using the original parameters of the model, is really $\alpha_\Delta=\alpha(\omega_c/\Delta)^{1-s}$. Following previous literature, we work with $\omega_c/\Delta=10$.

\begin{figure}
    \centering
    \includegraphics[width=0.98\linewidth]{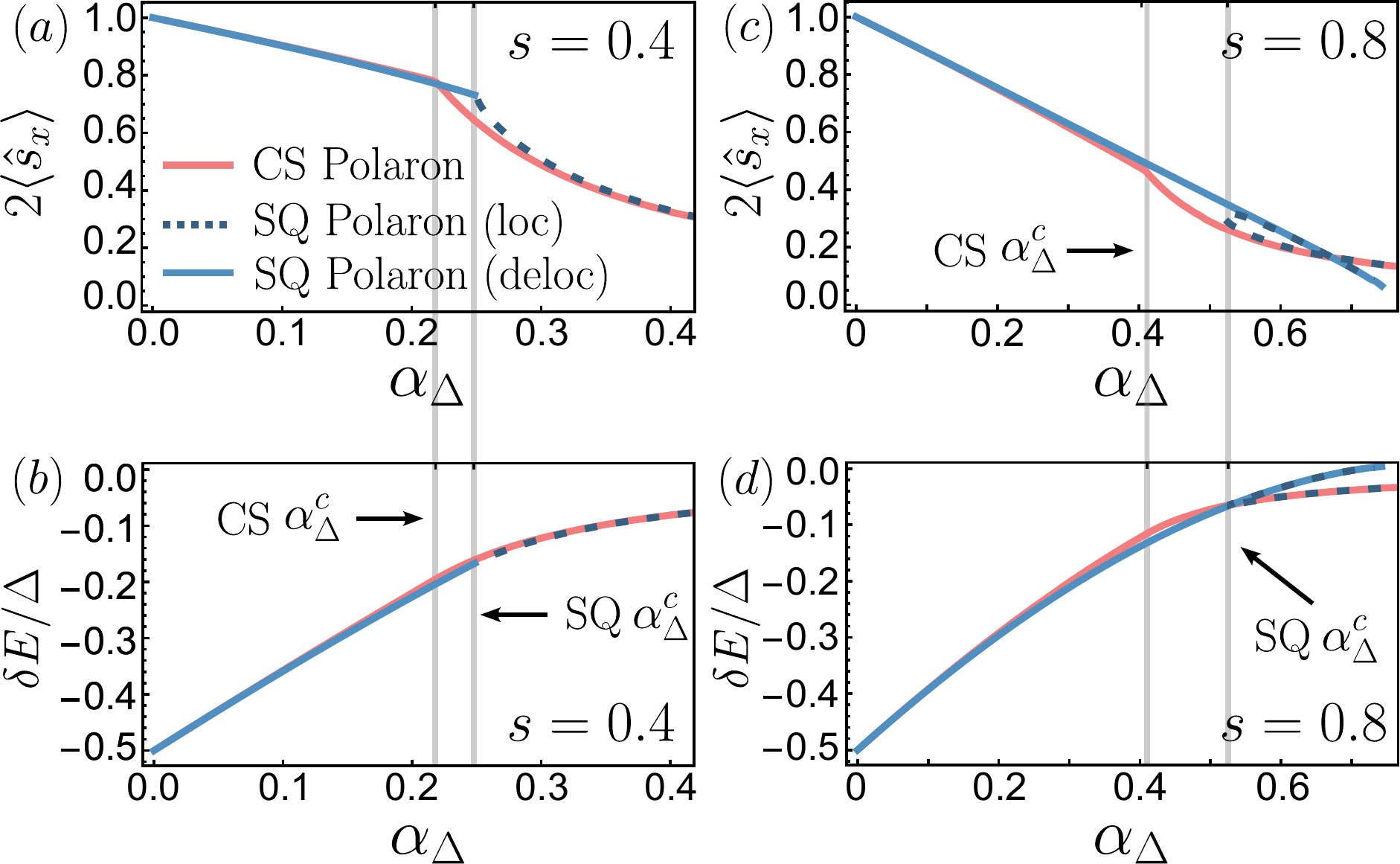}
    \caption{Coherence $\braket{\hat{s}_x}$ and ground state energy gain $\delta E/\Delta$ as a function of scaled coupling $\alpha_\Delta=\alpha(\omega_c/\Delta)^{1-s}$ for $s=0.4$ [(a),(b)] and $s=0.8$ [(c), (d)]. We show the results from Refs.~\cite{SilbeyHarris1984,Chin2011} at $\omega_c/\Delta=10$ in solid red (coherent state polaron, CS Polaron for short) and our results (squeezed polaron, SQ Polaron) in solid blue (delocalized branch) and dashed blue (localized branch). Vertical grey lines are the critical points $\alpha_\Delta^c$ predicted by the CS and SQ polarons.}
    \label{fig:Observables}
\end{figure}

\begin{figure}
    \centering
    \includegraphics[width=0.98\linewidth]{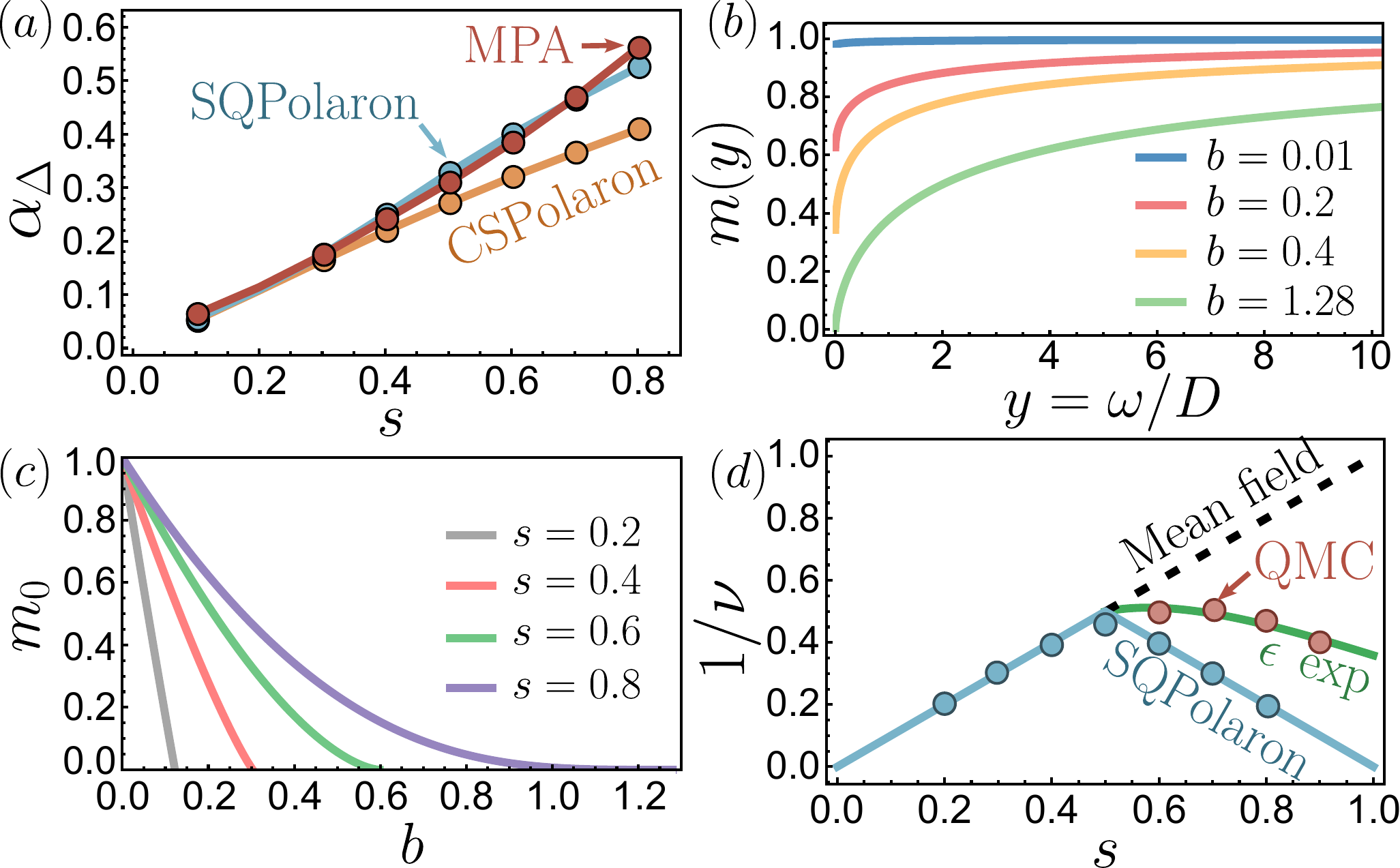}
    \caption{(a) Critical points $\alpha_\Delta^c$ predicted by the coherent state polaron~\cite{Chin2011} and the squeezed polaron vs. numerical results from a variational multipolaron ansatz (MPA)~\cite{SHEN20235} at $\Delta/\omega_c=0.1$. (b) Function $m(y)$ as a function of its argument $y$ ($y=\omega/D$ in the delocalized phase), for $s=0.8$ and $b=0.01,0.2,0.4$ and $b=b_c=1.28$. (c) Zero frequency value $m_0$ as a function of $b$ for different $s=0.2,0.4,0.6,0.8$. (d) Exponent $\nu$ as a function of $s$ of the squeezed polaron (blue dots are numerical), two loop $\epsilon$ expansion~\cite{Fisher1972} and Quantum Monte Carlo~\cite{Uzelac2001}.} 
    \label{fig:M(y)Properties}
\end{figure}

We solve numerically Eq.~(\ref{eqn:SqueezedPolaron:IntegralM(y)}) and Eq.~(\ref{eqn:SqueezedPolaron:Observables})~\cite{SM} and show in Fig.~\ref{fig:Observables} the energy gain $\delta E$ and coherence $\braket{\hat{s}_x}$ as a function of $\alpha_\Delta$ for $s=0.4,0.8$. We include the coherent state polaron and both branches of the squeezed polaron. For both values of $s$ we see small improvements in the variational energy, which is second-order sensitive to variations in the wavefunction. In contrast, changes in $\braket{\hat{s}_x}$ are comparatively larger, especially for $s=0.8$. For $s\lesssim 0.4$, the localized and delocalized branches are continuously connected. For larger $s$, the localized branch displays re-entrant behaviour and the minimal energy configuration jumps discontinuously between the delocalized and localized branches. The coupling at which the system switches between branches identifies the critical point and is shown in Fig.~\ref{fig:M(y)Properties}(a), compared against numerical results obtained via a Multiple Polaron Ansatz~\cite{SHEN20235}. Comparisons against other numerical works~\cite{Wong2008,Winter2009,2012Guo} can be found in the SM~\cite{SM}. The origin of the spurious first order jump is discussed later in the paper.

\noindent\textit{Delocalized branch:} In this branch $\theta=\alpha_\beta=0$ and $y_\beta=\omega_\beta/D$. Furthermore, the parameter $b$ is numerically equal to a renormalized coupling $\alpha_\Delta(\Delta/D)^{1-s}$.

All the physical properties of our wavefunction are a consequence of the structure of $m(y)$, as determined by Eq.~(\ref{eqn:SqueezedPolaron:IntegralM(y)}). In particular, the behaviour of $m(y)$ for small frequencies dominates the low frequency properties of the correlated displacements $q_\beta$ in Eq.~(\ref{eqn:SqueezedPolaron:Displacements}) and is thus relevant for the localization transition. Profiles for $m(y)$ as a function of $y=\omega/D$ at fixed $s=0.8$ and different $b$ are shown in Fig~\ref{fig:M(y)Properties}(b). In all cases, when $y\gg 1$ we have $m(y)\approx 1$. In the opposite regime of $y\ll 1$, $m(y)$ tends to a constant $m_0$. Hence, the displacements $q_\beta$ at small $\omega$ are finite and given by $g_\beta/(Dm_0)$.

The value of $m$ at zero frequency, $m_0$, decreases monotonically with increasing $b$ until it reaches $0$ at a value $b_c$ [see Fig.~\ref{fig:M(y)Properties}(c)]. When $s\lesssim 0.4$, the associated $\alpha_\Delta$ coincides with the critical point of the transition, but is in the higher energy branch when $s\gtrsim0.4$.

\noindent\textit{Delocalized branch near $b_c$:} At $b=b_c$, we have $m_c(y)\approx \pi b \cot(\pi s/2) y^s$ for small $y$~\cite{SM}. This determines the low frequency behaviour of the displacement and squeezing amplitudes~\cite{SM}
\begin{align}\begin{split}
    \big\langle\hat{s}_z(\hat{a}_\beta+\hat{a}_\beta^\dagger)\big\rangle&=\frac{q_\beta}{2}=\frac{g_\beta}{2\pi\alpha \omega_c^{1-s}\cot(\pi s/2)\omega_\beta^s}\\[5pt]
    \big\langle(\hat{a}_\beta+\hat{a}_\beta^\dagger)^2\big\rangle&=1+\frac{g_\beta^2(1+s)}{2\pi\alpha\omega_c^{1-s} \cot(\pi s/2)\, \omega_\beta^{1+s}}
\end{split}\end{align}
These results capture the correct frequency and $s$-dependence of these observables, obtained previously via analytical arguments based on diagrammatics~\cite{Florens2011,Blunden-Codd2017}. Furthermore, such scaling laws are in general incompatible with variational wavefunctions that involve a finite number of coherent state polarons~\cite{Blunden-Codd2017}.

As $m_0\to0$, the energy required to excite the boson cloud around the spin approaches $0$. This emerging scale $\omega_*$ is parametrically smaller than $D$ and governs the low frequency properties of the system. This is reflected in the function $m(y)$, which for frequencies $\omega\ll D$ acquires a scaling form~\cite{SM}
\begin{equation}
    m(y)\approx m_0\,r\left(\omega/\omega_*\right),
\end{equation}
with $r(0)=1$ and $\omega_*=D(m_0/b)^{1/s}\ll D$. This scaling function determines the physics of the system when $\omega\ll D$, and interpolates between the ``critical" regime $\omega\gg \omega_*$ and the ``non-critical" regime $\omega\ll \omega_*$. 

Analytical control over $r(\omega/\omega_*)$ grants us access to the relation $\omega^*\propto|\alpha_\Delta^c-\alpha_\Delta|^{\nu}$, which defines the critical exponent $\nu$. Using Eq.~(\ref{eqn:SqueezedPolaron:IntegralM(y)}) we find~\cite{SM} $\nu^{-1}=\text{min}(s,1-s)$, which is corroborated by numerical solution of Eq.~(\ref{eqn:SqueezedPolaron:IntegralM(y)}) [see Fig~\ref{fig:M(y)Properties}(d)]. For $s<0.5$ we recover the correct mean field exponent $\nu^{-1}=s$. For $s>0.5$ we find instead the non mean field value $\nu^{-1}=1-s$. Although not quantitatively accurate [see Fig.~\ref{fig:M(y)Properties}(d)], and in a  regime where the localized branch is energetically favored, the squeezed polaron captures the decreasing trend of $\nu^{-1}$ with increasing $s$.

\noindent\textit{Scattering:} With an analytical description of the boson bath at hand, we can now consider the effects of probing the environment. For instance, the spin-boson model can be re-interpreted as the long-range interaction between a spin 1/2 impurity and a boson~\cite{Bera2016} propagating on the half-line. Then $\beta\to k>0$ can be taken to be a momentum label, associated to a normal mode with mode function $\propto \cos(kx)$, where $x>0$ is the position on a line of length $L$. The dispersion is linear $\omega_k=k$ and the couplings take the form $g_k^2=2\pi\times2\alpha\omega_c^{1-s}k^s/L$. 

\begin{figure}
    \centering
    \includegraphics[width=0.98\linewidth]{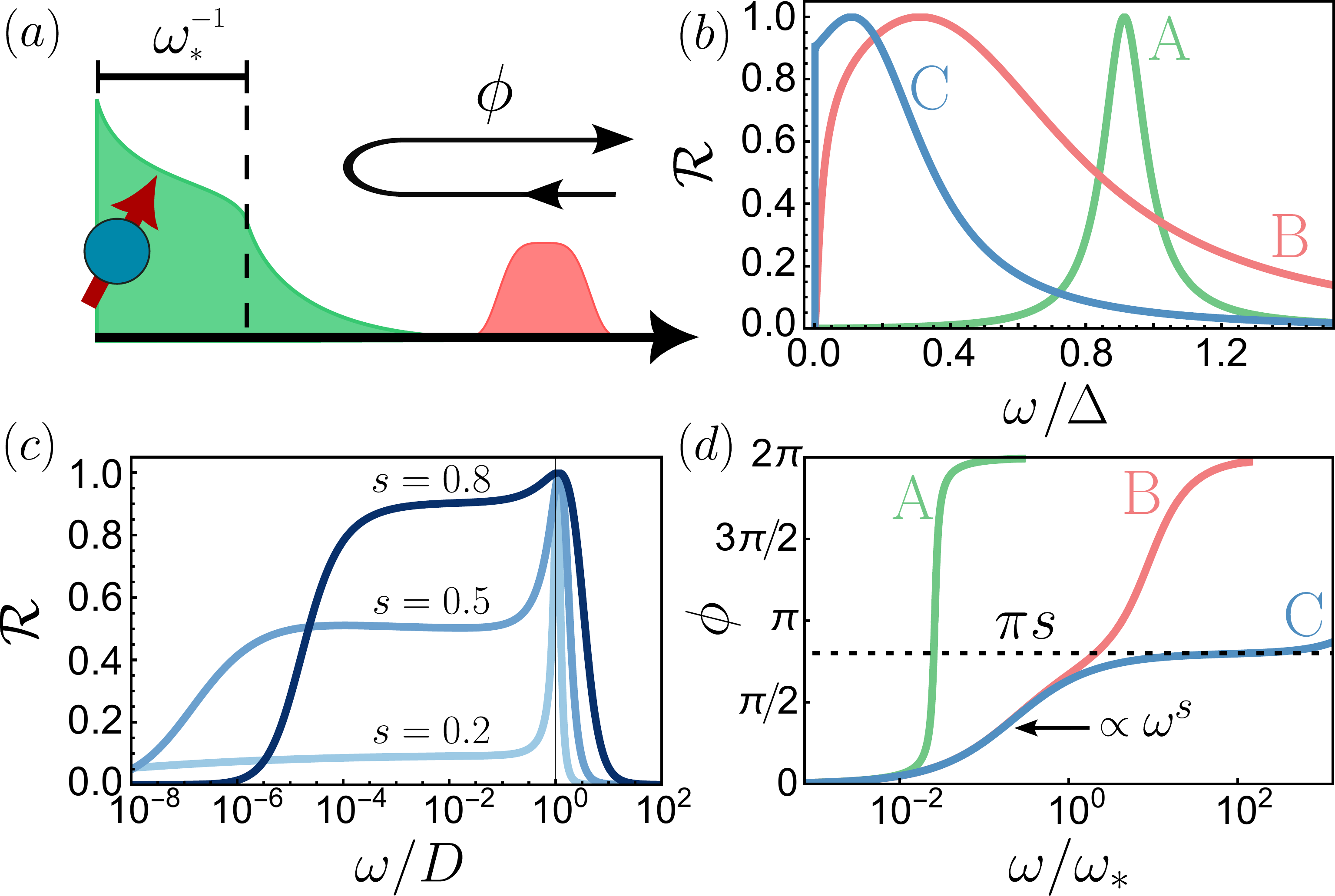}
    \caption{(a) Scattering of bosons in the half-line. (b) We show the full-line reflection coefficient $\mathcal{R}=\sin(\phi/2)^2$ as a function of $\omega/\Delta$ for $s=0.8$ and $\alpha_\Delta=0.05,0.5$ and $0.7$, denoted A, B and C, respectively. These correspond to $m_0=0.9,0.16, 0.002$ and $\omega_*/\Delta=35,0.07,5\times 10^{-5}$. (c) $\mathcal{R}$ for $s=0.2,0.6,0.8$ near criticality, all with $\omega_*/D=5\times 10^{-6}$ (corresponding to $\alpha_\Delta=0.121,0.4297,1.15$). (d) Phase shift $\phi$ at (low) frequencies comparable to $\omega_*$ for same parameters as in (b). }
    \label{fig:Scattering}
\end{figure}

In the delocalized phase, there is a boson cloud of size $\sim (\omega_*)^{-1}$ around the impurity spin, which can be probed via scattering~\cite{Zarand2003,Borda2007}, see Fig.~\ref{fig:Scattering}(a). The effective Hamiltonian describing scattering of a single photon is~\cite{SM}
\begin{equation}\label{eqn:effHamiScattering}
    \hat{h}_{\text{eff}}=\hat{h}_{\text{cloud}}+\frac{D}{2}(1+\hat{s}_x)+\hat{B}\sigma^+_x+\hat{B}^\dagger\hat{\sigma}_x^-,
\end{equation}
where $\hat{h}_{\text{cloud}}=\sum_k\omega_k\hat{a}_k^\dagger\hat{a}_k+D(\sum_{k}q_k\hat{p}_k)^2/2$ defines photons dressed by the impurity cloud, $\hat{\sigma}^{\pm}_x=\hat{s}_y\pm i\hat{s}_z$ create/destroy spin excitations along the $x$ direction and $\hat{B}=D\sum_k q_k[\hat{p}_k-im(y_k)\hat{x}_k]/\sqrt{2}$ annihilates dressed photons. The effective $\hat{h}_{\text{eff}}$ is obtained by acting with $\hat{\mathcal{D}}_c$ on $\hat{H}$ in Eq.~(\ref{eqn:Model:Hamiltonian}), normal-ordering with respect to the variational ground state $\ket{\phi}$ and keeping only terms that are up to quadratic in boson operators and conserve the number of dressed (as opposed to bare~\cite{Shi2018}) excitations.

In the half-line, a single photon with frequency $\omega$ will acquire a phase shift $\phi(\omega)$. In Fig.~\ref{fig:Scattering}(b) we plot $\mathcal{R}=\sin(\phi/2)^2$, which is equal to the reflection coefficient in the full-line, for $s=0.8$.  At weak coupling ($\alpha_\Delta\ll\alpha_\Delta^c$, green line, A), $\mathcal{R}$ displays a typical Lorentzian resonance centered around a shifted spin flip frequency $\omega \sim D$. At larger $\alpha_\Delta$ (red line, B), the peak gets broadened, shifted and distorted. This trend continues as $\alpha^c\to \alpha_\Delta^c$ (blue line, C),  but there is now a rapid decrease in $\mathcal{R}$ at very low frequencies. This is more easily seen in a logarithmic scale [Fig.~\ref{fig:Scattering}(b)], which also shows (for $s=0.2,0.5,0.8$) that there is broad frequency plateau in which $\mathcal{R}$ is constant. The value of $\mathcal{R}$ at the plateau increases with $s$.

The plateau appears in the critical regime, $\omega_*\ll\omega\ll D$, which spans a wide frequency range when $\alpha_\Delta\approx\alpha_\Delta^c$. In this regime, we have constant $\phi_c=\pi s$~\cite{SM} and hence fixed $\mathcal{R}_c=\sin(\pi s/2)^2$. At smaller frequencies, $\omega\ll \omega_*$, the photon does not have enough energy to excite the bosonic cloud, and $\phi \approx 2\pi(\omega/\omega_*)^s$~\cite{SM} is small. In the half-line, such shifts could be probed via interference with a reference wave using a homodyne detection setup. When $d\phi/d\omega\neq 0$, the wave also disperses, thus introducing a measurable time-delay and wave-packet distortion.

We now discuss the applicability of these results. When $\omega\ll\omega_*$ the physics is controlled by the energy gap, so we expect our results to be quantitatively accurate. In the critical regime, the physics is controlled by a nonlocal CFT~\cite{Benedetti2025,Benedetti2026}, from which we can derive that the $1\to1$ $S$-matrix has the form $S(E)=1+iC'e^{i\pi  s/2}$, with $C'>0$ and frequency independent~\cite{SM}. In particular, adding the requirement of no multiparticle production fixes $C'=2\sin(\pi s/2)\to S(E)=e^{i\pi s}$, reducing to our variational result. Since the associated CFT is gaussian when $s<0.5$, we expect our low energy predictions to be quantitatively accurate. For $s>0.5$, multiparticle production will modify $\mathcal{R}_c$.

\noindent\textit{Localized branch:} In the localized branch Eq.~(\ref{eqn:SqueezedPolaron:Angle}) has two non-zero equal energy solutions, and the parameter $b$ now equals $\alpha_\Delta(\Delta/D)^{1-s}(\cos\theta)^{3-s}$. As discussed in Ref.~\cite{Chin2011}, in the localized branch the $\alpha_\beta$ diverge as $\omega\to 0$ [see Eq.~(\ref{eqn:SqueezedPolaron:Displacements})] leading to an effective $z$ field for the spin in Eq.~(\ref{eqn:Model:Hamiltonian}) that tilts it away from the $-x$ axis. 

We can recast Eq.~(\ref{eqn:SqueezedPolaron:Angle}) as~\cite{SM} $m_0=(\sin\theta)^2$ using Eq.~(\ref{eqn:SqueezedPolaron:IntegralM(y)}). Thus, the identification of the critical point with $m_0=0$ (valid for $s\lesssim 0.4$) is consistent with a continuous variation of the angle $\theta$ away from $0$. We can also use this relation to extract the critical exponent $\beta$. In the region $s<0.5$, we find again the correct mean field value $\beta=1/2$. When $0.5<s<1$ the squeezed polaron predicts a non mean-field $\beta=0.5s/(1-s)$.

We believe that the spurious first-order jump between localized and delocalized branches in Fig.~\ref{fig:Observables} is caused by an inadequate handling of fluctuations near the critical point. The true wavefunction in the localized phase is more accurately taken to be a cat-state superposition of both localized solutions~\cite{Zheng2013,Zheng2026}. Hence, in the delocalized phase but near criticality there should also be cat-like fluctuations, but such that the phase space separation between cat lobes is not completely resolvable within the quantum noise of the squeezed polaron. 

\textit{Summary and conclusions:}
In this paper we analyzed the consequences of adding Gaussian fluctuations to the standard polaron variational wavefunction of the spin-boson model. By including boson-boson correlations we found a correct qualitative (and in some cases quantitative) description of most properties of the spin-boson ground state. These include the emergence and nature of the critical fluctuations that arise in the boson sector as the localization phase transition is approached, the switch from mean-field to non mean-field behaviour as the bath exponent $s$ is increased, and the effects of scattering bosons off the spin impurity. 

Extensions of this work will assess the effect of adding Gaussian fluctuations to bipolaron ansatzes~\cite{Bera2014b,Bera2014} as a way of improving the properties of the wavefunction near the transition~\cite{Zheng2026}, studying in more detail the implications for the Ohmic case and addressing the case of finite temperature. Furthermore, it would be useful to get a similar analytical grasp of the bosonic sector in the dynamics of the spin-boson model~\cite{Nalbach2013}, and use this information to design feedback operations with the objective of enhancing spin-coherence.

\begin{acknowledgments}
{\it Acknowledgements:} 
We thank Alex Chin for helpful discussions. This work was supported by the Simons Investigator Award (Grant No. 511029) and the Engineering and Physical Sciences Research Council [grant numbers EP/V062654/1 and EP/Y01510X/1].
\end{acknowledgments}
\bibliography{library}

\balancecolsandclearpage
\onecolumngrid
\appendix

\setcounter{equation}{0}
\setcounter{figure}{0}
\setcounter{page}{1}
\renewcommand{\theequation}{S\arabic{equation}}
\renewcommand{\thefigure}{S\arabic{figure}}
\renewcommand{\thepage}{S\arabic{page}}
\renewcommand{\bibnumfmt}[1]{[S#1]}
\renewcommand{\citenumfont}[1]{S#1}

\section*{Appendix Contents}

\noindent Appendix A: \hyperref[appA]{Minimization of energy functional} \dotfill \pageref{appA} \\
\noindent Appendix B: \hyperref[appB]{Correlations of Gaussian state} \dotfill \pageref{appB} \\
\noindent Appendix C: \hyperref[appC]{Numerical solution of integral equation} \dotfill \pageref{appC}\\
\noindent Appendix D: \hyperref[appD]{Critical coupling comparisons} \dotfill \pageref{appD} \\
\noindent Appendix E: \hyperref[appE]{Low frequency behaviour of $m(y)$ at the critical point} \dotfill \pageref{appE} \\
\noindent Appendix F: \hyperref[appF]{Scaling behaviour of $m(y)$ at low frequency} \dotfill \pageref{appF} \\
\noindent Appendix G: \hyperref[appG]{Correlation length exponent} \dotfill \pageref{appG} \\
\noindent Appendix H: \hyperref[appH]{Effective Hamiltonian for scattering} \dotfill \pageref{appH} \\
\noindent Appendix I: \hyperref[appI]{Scattering using CFT methods} \dotfill \pageref{appI} \\
\noindent Appendix J: \hyperref[appJ]{Relation between spin orientation and $m_0$} \dotfill \pageref{appJ} \\

\vspace{1cm}

\section{MINIMIZATION OF ENERGY FUNCTIONAL}\label{appA}
Here we perform the minimization of the Hamiltonian in the dressed frame with respect to the manifold of states $\ket{\psi}=\mathcal{D}_c\mathcal{D}_u\hat{R}_y(\theta)\ket{-x}\ket{\phi}$, where $\ket{-x}$ is a spin state pointing in the $-x$ direction, $\hat{R}_y(\theta)=\exp(-i\theta\hat{s}_y)$ is a rotation about $+y$  by an angle $\theta$, $\mathcal{D}_u=\exp\big[\sum_\beta \alpha_\beta(\hat{a}_\beta^\dagger-\hat{a}_\beta)\big]$ is a bosonic displacement operator ($\mathcal{D}^\dagger_u\hat{x}_\beta\mathcal{D}_u=\hat{x}_\beta+\sqrt{2}\alpha_\beta$), $\mathcal{D}_c=\exp\big[-\hat{s}_z\sum_\beta q_\beta(\hat{a}_\beta^\dagger-\hat{a}_\beta)\big]$ and $\ket{\phi}$ is a zero mean gaussian state. For computations, it's convenient to apply the correlated displacements directly on the Hamiltonian since the remaining part does not have spin-boson correlations:
\begin{align}\begin{split}\label{eqn:appendix:CorrelatedHamiltonian}
    \hat{H}_c=\mathcal{D}_c^\dagger\hat{H}\mathcal{D}_c&=\frac{\Delta}{2}\left[\hat{\sigma}_+e^{-i\sum_\beta q_\beta(\hat{a}_\beta-\hat{a}_\beta^\dagger)}+\hat{\sigma}_- e^{i\sum_\beta q_\beta(\hat{a}_\beta-\hat{a}_\beta^\dagger)}\right]+\sum_\beta(q_\beta-\omega_\beta q_\beta)\hat{s}_z(\hat{a}_\beta+\hat{a}_\beta^\dagger)+\sum_\beta \omega_\beta\hat{a}^\dagger_\beta\hat{a}_\beta\\
    &+\sum_\beta \frac{(q_\beta-\omega_\beta q_\beta)^2}{4\omega_\beta}-\sum_\beta\frac{g_\beta^2}{4\omega_\beta}
\end{split}\end{align}
Now projecting onto the rest of the state and using the gaussianity property of $\ket{\phi}$ leads to the following function of $q_\beta,\alpha_\beta,\ket{\phi},\theta$
\begin{align}\begin{split}
    E&=-\frac{\Delta\cos\theta}{2}\exp\left(-\sum_{\beta,\beta'}q_\beta q_{\beta'}\braket{\hat{p}_\beta\hat{p}_{\beta'}}_\phi\right)+\frac{1}{2}\sum_\beta\omega_\beta\langle\hat{x}_\beta^2+\hat{p}_\beta^2-1\rangle_\phi\\
    &+\sum_\beta (g_\beta-\omega_\beta q_\beta)\alpha_\beta\sin\theta+\sum_\beta \omega_\beta \alpha_\beta^2+\sum_\beta\frac{(g_\beta-\omega_\beta q_\beta)^2}{4\omega_\beta}\underbrace{-\sum_\beta\frac{g_\beta^2}{4\omega_\beta}}_{E_{\text{ad}}},
\end{split}\end{align}
where $\hat{x}_\beta=(\hat{a}_\beta+\hat{a}_\beta^\dagger)/\sqrt{2}$ and $\hat{p}_\beta=-i(\hat{a}_\beta-\hat{a}_\beta^\dagger)/\sqrt{2}$ are quadratures. For simplicity, we also define the renormalized tunneling rate 
\begin{equation}
    D=\Delta\exp\left(-\sum_{\beta,\beta'}q_\beta q_{\beta'}\braket{\hat{p}_\beta\hat{p}_{\beta'}}_\phi\right).
\end{equation}
Minimizing with respect to the state $\ket{\phi}$ subject to the constraint $\braket{\phi|\phi}=1$ leads to the gaussian eigenvalue equation
\begin{equation}\label{eqn:appB_gauss}
    \boxed{\left[\frac{\hat{X}^T\Omega\hat{X}}{2}+\frac{\hat{P}^T(\Omega+D\cos\theta qq^T)\hat{P}}{2}\right]\ket{\phi}=\mu\ket{\phi},}
\end{equation}
where $\hat{X}=(\hat{x}_\beta),\,\hat{P}=(\hat{p}_\beta),\,q=(q_\beta)$ are vectors, $\Omega=\text{diag}(\omega_\beta)$ is a diagonal matrix and $\mu$ is a Lagrange multiplier. Minimization with respect to $q_\beta$ and $\alpha_\beta$ (expressing $q_\beta,\,\alpha_\beta$ in vectorized form) leads to the equations
\begin{align}\label{eqn:appB_vectors}
    \begin{split}
        \boxed{D\cos\theta\braket{\hat{P}\hat{P}^T}q-\Omega \alpha\sin\theta+\frac{\Omega q-g}{2}=0}&\\
        \boxed{\frac{\sin\theta\,(g-\Omega q)}{2}+\Omega \alpha=0}&
    \end{split}
\end{align}
Finally, minimization with respect to $\theta$ leads to
\begin{equation}\label{eqn:appB_theta}
    \boxed{\frac{D\sin\theta}{2}+\cos\theta \,\alpha^T(g-\Omega q)=0.}
\end{equation}
\subsection{Boson correlations and displacements}
We can actually proceed with the solution without using the equation coming from minimization over $\theta$, very much in the spirit of Ref.~\cite{Chin2011}, and then adding the constraint coming from minimizing over $\theta$ in the end. From Eq.~(\ref{eqn:appB_vectors}) we can express $q$ in terms of $g$ as
\begin{align}\label{eqn:appB_q1}
    \begin{split}
        \left(\cos\theta\,\Omega+2D\braket{\hat{P}\hat{P}^T}\right)q&=g\cos\theta
    \end{split}
\end{align}
The correlator $\braket{\hat{P}\hat{P}^T}$ can be written down in matrix form using Eq.~(\ref{eqn:appB_gauss}):
\begin{equation}
    \braket{\hat{P}\hat{P}^T}=\frac{1}{2}\Omega^{1/2}\frac{1}{\sqrt{\Omega^2+D\cos\theta\,\Omega^{1/2}qq^T\Omega^{1/2}}}\Omega^{1/2}
\end{equation}
Since the matrix inside the square root is a diagonal matrix plus a rank 1 perturbation, we can make progress by using resolvent methods. We don't actually need the full matrix, just its action of the vector $q$, which ends up being simpler. After some manipulations (detailed later in this Supplementary Material) we get
\begin{equation}\label{eqn:AppB_PPq}
    \braket{\hat{P}\hat{P}^T}q=\frac{1}{2}\left\{\int_{\mathcal{C}}\frac{dz}{2\pi i\sqrt{z}}\frac{\Omega}{z-\Omega^2}\frac{1}{1-D\cos\theta\,\left[q^T\left(\frac{\Omega}{z-\Omega^2}\right)q\right]}\right\}q,
\end{equation}
where $\mathcal{C}$ goes from $\infty+i 0^+$ to $0$ and then from $0$ to $\infty-i0^+$. Note that the object in the curly brackets can be written as the matrix function $l(\Omega)$, where $l$ is the function defined by
\begin{equation}\label{eqn:appB_l1}
    l(x)=\int_{\mathcal{C}}\frac{dz}{2\pi i\sqrt{z}}\frac{x}{z-x^2}\frac{1}{1-D\cos\theta\,\left[q^T\left(\frac{\Omega}{z-\Omega^2}\right)q\right]}.
\end{equation}
Then, from Eq.~(\ref{eqn:appB_q1}) we get
\begin{equation}
    q=\left[\frac{\cos\theta}{\Omega\cos\theta+D\,l(\Omega)}\right]g
\end{equation}
Plugging this form back into Eq.~(\ref{eqn:appB_l1}) we get a self-consistency condition for the function $l$
\begin{equation}\label{eqn:appB_l2}
    l(x)=\int_{\mathcal{C}}\frac{dz}{2\pi i\sqrt{z}}\frac{x}{z-x^2}\frac{1}{1-D(\cos\theta)^3\,\left[g^T\left(\frac{\Omega}{z-\Omega^2}\right)\left(\frac{1}{\Omega\cos\theta+Dl(\Omega)}\right)^2g\right]}.
\end{equation}
In the continuum limit, this becomes
\begin{equation}\label{eqn:appB_lcontinuum}
    l(x)=\int_{\mathcal{C}}\frac{dz}{2\pi i\sqrt{z}}\frac{x}{z-x^2}\left\{1-2\alpha \omega_c^{1-s}D(\cos\theta)^3\,\int_0^{\omega_c}\frac{u^{s+1}du}{(z-u^2)\big[u \cos\theta+Dl(u)\big]^2}\right\}^{-1}.
\end{equation}
Introducing the function $m(\omega\cos\theta/D)=l(\omega)$ and changing variables appropriately (both in the $z$ and $u$ integrals), we get
\begin{equation}
    m(y)=\int_{\mathcal{C}}\frac{d\tilde{z}}{2\pi i\sqrt{\tilde{z}}}\frac{y}{\tilde{z}-y^2}\left\{1-2\alpha \omega_c^{1-s}D^{1-s}(\cos\theta)^{3-s}\,\int_0^{\omega_c\cos\theta/D}\frac{v^{s+1}dv}{(\tilde{z}-v^2)\big[v+m(v)\big]^2}\right\}^{-1}.
\end{equation}
Finally, we deform the contour from the branch cut at positive $z$ to the negative $z$ axis, getting contributions from the square root branch cut. After changing variables to $z_*=\sqrt{\tilde{z}}$ and defining $\alpha_D=\alpha(\omega_c/D)^{1-s}$, we get
\begin{equation}\label{eqn:AppB_mfinal}
    m(y)=\int_0^{\infty}\frac{2ydz_*}{\pi(y^2+z_*^2)}\frac{1}{1+2\alpha_D\cos\theta^{3-s}\int_0^{\omega_c\cos\theta/D}\frac{v^{s+1}dv}{(v^2+z_*^2)\big[v+m(v)\big]^2}},
\end{equation}
which is Eq.~(\ref{eqn:SqueezedPolaron:IntegralM(y)}) in the main text after defining $b=\alpha_D(\cos\theta)^{3-s}$ and $y_c=\omega_c\cos\theta/D$. Thus we have
\begin{align}\label{eqn:AppB_qvector}
    \begin{split}
        q=\left[\frac{\cos\theta}{\Omega \cos\theta+D m\left(\frac{\Omega\cos\theta}{D}\right)}\right]g,
    \end{split}
\end{align}
which is Eq.~(\ref{eqn:SqueezedPolaron:Displacements}) component by component if we defined $y=\Omega\cos\theta/D$. Similarly,
\begin{equation}
    \alpha=-\frac{\sin\theta}{2\Omega}\left[\frac{D m\left(\frac{\Omega\cos\theta}{D}\right)}{\Omega\cos\theta+D m\left(\frac{\Omega\cos\theta}{D}\right)}\right]g.
\end{equation}
The $\alpha_\beta$ go to $0$ when $\theta\to 0$, as they should.
\subsubsection{Observables}
From the definition of $D$, we have
\begin{equation}
    D=\Delta\exp\left(-q^T\braket{\hat{P}\hat{P}^T}q\right)=\Delta\exp\left[-\frac{1}{2}q^T l(\Omega)q\right]=\Delta\exp\left\{-\frac{1}{2}g^T\left[ \frac{(\cos\theta)^2l(\Omega)}{\big(\Omega\cos\theta+Dl(\Omega)\big)^2}\right]g\right\}.
\end{equation}
In the continuum limit, we thus have
\begin{equation}
    D=\Delta\exp\left\{-\alpha\omega_c^{1-s}(\cos\theta)^2\int_0^{\omega_c}\frac{l(u)u^sdu}{\big[u\cos\theta+Dl(u)\big]^2}\right\}
\end{equation}
Changing variables we get
\begin{equation}
    D=\Delta\exp\left\{-\alpha_D(\cos\theta^{1-s})\int_0^{\omega_c\cos\theta/D}\frac{m(v)v^sdv}{\big[v+m(v)\big]^2}\right\},
\end{equation}
recovering Eq.~(\ref{eqn:SqueezedPolaron:Observables}), again after identifying $b=\alpha_D(\cos\theta)^{3-s}$. The equation for the bare coupling $\alpha_\Delta=\alpha(\omega_c/\Delta)^{1-s}$ is obtained from the previous one by taking its $(s-1)$-th power
\begin{equation}
    \alpha_\Delta=\alpha_D \exp\left\{-(1-s)\alpha_D(\cos\theta^{1-s})\int_0^{\omega_c\cos\theta/D}\frac{m(v)v^sdv}{\big[v+m(v)\big]^2}\right\}
\end{equation}

For the energy we replace the values of $\alpha_\beta$, and rewrite it in a convenient way
\begin{align}\begin{split}
    E&=-\frac{D\cos\theta}{2}+(\cos\theta)^2\sum_\beta\frac{(g_\beta-\omega_\beta q_\beta)^2}{4\omega_\beta}+\sum_\beta\left(\frac{\omega_\beta}{2}\braket{\hat{x}_\beta^2}+\frac{\omega_\beta}{2}\braket{\hat{p}_\beta^2}-\frac{\omega_\beta}{2}\right)+\frac{D\cos\theta}{2}\sum_{\beta \beta'}q_\beta q_{\beta'}\braket{\hat{p}_\beta\hat{p}_{\beta'}}\\
    &-\frac{D\cos\theta}{2}\sum_{\beta\beta'}q_\beta q_{\beta'}\braket{\hat{p}_\beta\hat{p}_{\beta'}}-\frac{\omega_c\alpha}{2s}
\end{split}\end{align}
Plugging the solution for $q_\beta$ yields
\begin{align}
    \begin{split}
         E&=-\frac{D\cos\theta}{2}+\frac{D^2(\cos\theta)^2}{4}g^T\left\{\frac{\left[m\left(\frac{\Omega\cos\theta}{D}\right)\right]^2}{\Omega\left[\Omega\cos\theta+Dm\left(\frac{\Omega\cos\theta}{D}\right)\right]^2}\right\}g+\frac{1}{2}\mathrm{Tr}\left\{\sqrt{\Omega^2+D\cos\theta qq^T}-\Omega\right\}\\
    &-\frac{D\cos\theta}{2}q^T\braket{\hat{P}\hat{P^T}}q-\frac{\omega_c\alpha}{2s}.
    \end{split}
\end{align}
Changing variables and calculating the trace via resolvent methods leads to
\begin{align}
    \begin{split}
        E&=-\frac{D\cos\theta}{2}+\frac{Db}{2\cos\theta}\int_0^{y_c}\frac{y^{s-1}m(y)^2dy}{\big[y+m(y)\big]^2}-\frac{Db}{2\cos\theta}\int_0^{y_c}\frac{y^s m(y)dy}{\big[y+m(y)\big]^2}+\frac{Db}{2\cos\theta}\int_0^{y_c}\frac{y^{s}\partial_y\big[y m(y)\big]dy}{\big[y+m(y)\big]^2}-\frac{\omega_c\alpha}{2s}\\
    &=-\frac{D\cos\theta}{2}+\frac{Db}{2\cos\theta}\int_0^{y_c}\frac{y^{s-1}m(y)^2dy}{\big[y+m(y)\big]^2}+\frac{Db}{2\cos\theta}\int_0^{y_c}\frac{y^{s+1}\partial_y\big[ m(y)\big]dy}{\big[y+m(y)\big]^2}-\frac{\omega_c\alpha}{2s}
    \end{split}
\end{align}
After using the relation
\begin{equation}
    \int_0^{y_c}\frac{y^{s+1}\partial_y m(y)dy}{\big[y+m(y)\big]^2}=\frac{y_c^s}{y_c+m(y_c)}-(s+1)\int_0^{y_c}\frac{y^{s-1}m(y)dy}{y+m(y)}+\int_0^{y_c}\frac{y^{s-1}m(y)^2 dy}{\big[y+m(y)\big]^2}+2\int_0^{y_c}\frac{y^sm(y)dy}{\big[y+m(y)\big]^2},
\end{equation}
the energy becomes
\begin{equation}
    E=-\frac{D\cos\theta}{2}+\frac{Db(1-s)}{2\cos\theta}\int_0^{y_c}\frac{y^{s-1}m(y)dy}{y+m(y)}-\frac{\omega_c\alpha}{2s}+\frac{Db}{2\cos\theta}\left[\frac{y_c^s}{y_c+m(y_c)}\right].
\end{equation}
The last term goes to $0$ when $y_c\to \infty$.
\subsection{Angle minimization}
Plugging in all the previous expressions into Eq.~(\ref{eqn:appB_theta}) leads to
\begin{equation}
    \frac{D\sin\theta}{2}=\frac{\cos\theta\sin\theta}{2}g^T\left[\frac{D^2m\left(\frac{\Omega\cos\theta}{D}\right)^2}{\Omega\big(\Omega\cos\theta+m\left(\frac{\Omega\cos\theta}{D}\right)\big)^2}\right]g.
\end{equation}
This has as solution either $\theta=0$ or (in the continuum limit)
\begin{equation}
    D=2\alpha\omega_c^{1-s}\cos\theta\int_0^{\omega_c}\frac{D^2m(u\cos\theta/D)^2u^{s-1}du}{\big[u\cos\theta+Dm(u\cos\theta/D)\big]^2}.
\end{equation}
After changing variables we thus get
\begin{equation}
    1=2\alpha_D(\cos\theta)^{1-s}\int_0^{\omega_c \cos\theta/D}\frac{v^{s-1}m(v)^2\,dv}{\big[v+m(v)\big]^2}.
\end{equation}
From Eq.~(\ref{eqn:AppB_mfinal}) we find it convenient to parameterize the solution in terms of $b=\alpha_D(\cos\theta)^{3-s}$. Then we have
\begin{align}\begin{split}
m(y)&=\int_0^{\infty}\frac{2ydz_*}{\pi(y^2+z_*^2)}\frac{1}{1+2b\int_0^{y_c}\frac{v^{s+1}dv}{(v^2+z_*^2)\big[v+m(v)\big]^2}}\\[5pt]
    \cos\theta^2&=2b\int_0^{y_c}\frac{v^{s-1}m(v)^2\,dv}{\big[v+m(v)\big]^2}\\[5pt]
     D&=\Delta\exp\left\{-\frac{b}{(\cos\theta)^2}\int_0^{y_c}\frac{m(v)v^sdv}{\big[v+m(v)\big]^2}\right\},\\[5pt]
     \alpha_\Delta&=\frac{b}{(\cos\theta)^{3-s}}\exp\left\{-\frac{(1-s)b}{(\cos\theta)^2}\int_0^{y_c}\frac{m(v)v^sdv}{\big[v+m(v)\big]^2}\right\},\\[5pt]
     E&=-\frac{D\cos\theta}{2}\left[1-\frac{b(1-s)}{(\cos\theta)^2}\int_0^{y_c}\frac{v^{s-1}m(v)dv}{v+m(v)}\right]+\frac{Db}{2\cos\theta}\left[\frac{y_c^s}{y_c+m(y_c)}\right]-\frac{\omega_c \alpha}{2s}
\end{split}\end{align}

\section{CORRELATIONS OF GAUSSIAN STATE}\label{appB}

Here we solve the gaussian problem defined by Eq.~(\ref{eqn:appB_gauss}), which we copy here for completeness
\begin{equation}
    \left[\frac{\hat{X}^T\Omega\hat{X}}{2}+\frac{\hat{P}^T(\Omega+\eta\, q q^T)\hat{P}}{2}\right]\ket{\phi}=\mu\ket{\phi}.
\end{equation}
where $\Omega$ is a diagonal matrix, $q$ is a vector of numbers, $\eta=D\cos\theta$ is a parameter and $\hat{X},\hat{P}$ are vectors of conjugate quadratures. Because this is a gaussian problem, the correlators can be written down directly
\begin{align}
    \begin{split}
        \braket{\hat{X}\hat{X}^T}&=\frac{1}{2}\Omega^{-1/2}\left(\Omega^2+\eta\, \Omega^{1/2}q q^T\Omega^{1/2}\right)^{1/2}\Omega^{-1/2}\\[5pt]
        \braket{\hat{P}\hat{P}^T}&=\frac{1}{2}\Omega^{1/2}\left(\Omega^2+\eta\, \Omega^{1/2}q q^T\Omega^{1/2}\right)^{-1/2}\Omega^{1/2}
    \end{split}
\end{align}
We will work with $\braket{\hat{P}\hat{P^T}}$. Similar manipulations should work for $\braket{\hat{X}\hat{X}^T}$. We first express the square root in terms of the resolvent
\begin{equation}
    \braket{\hat{P}\hat{P}^T}=\frac{1}{2}\Omega^{1/2}\int_{\mathcal{C}}\frac{dz}{2\pi i\sqrt{z}}\frac{1}{z-(\Omega^2+\eta\Omega^{1/2}qq^T\Omega^{1/2})}\Omega^{1/2},
\end{equation}
where $\mathcal{C}$ encircles all the eigenvalues of $\Omega^2+\eta\,\Omega^{1/2}qq^T\Omega^{1/2}$. We can guarantee that this is always the case by letting the contour run from $\infty+i\epsilon$ to $0$ and then from $0$ to $\infty-i\epsilon$. Because $qq^T$ is single-rank, we have then that
\begin{align}\begin{split}
    \frac{1}{z-\Omega^2-\eta\Omega^{1/2}qq^T\Omega^{1/2}}&=\left[\frac{1}{1-\frac{\eta}{z-\Omega^2}\Omega^{1/2}q q^T\Omega^{1/2}}\right]\frac{1}{z-\Omega^2}\\[5pt]
    &=\left[1+\frac{\eta}{z-\Omega^2}\Omega^{1/2}qq^T\Omega^{1/2}\times\frac{1}{1-\eta\,\left(q^T\frac{\Omega}{z-\Omega^2}q\right)}\right]\frac{1}{z-\Omega^2}
\end{split}\end{align}
Thus, we get
\begin{equation}
    \braket{\hat{P}\hat{P}^T}=\frac{1}{2}\int_{\mathcal{C}}\frac{dz}{2\pi i\sqrt{z}}\left[\frac{\Omega}{z-\Omega^2}+\frac{\eta}{1-\eta\left(q^T\frac{\Omega}{z-\Omega^2}q\right)}\frac{\Omega}{z-\Omega^2}qq^T\frac{\Omega}{z-\Omega^2}\right]
\end{equation}
In particular,
\begin{equation}
    \braket{\hat{P}\hat{P}^T}q=\frac{1}{2}\int_{\mathcal{C}}\frac{dz}{2\pi i\sqrt{z}}\frac{\Omega}{z-\Omega^2}\frac{1}{1-\eta \left(q^T\frac{\Omega}{z-\Omega^2}q\right)},
\end{equation}
which is Eq.~(\ref{eqn:AppB_PPq}) with $\eta=D\cos\theta$. We can also read-off
\begin{align}\begin{split}
    \braket{\hat{p}_\beta\hat{p}_{\gamma}}&=\frac{\delta_{\beta\gamma}}{2}+\frac{\eta \,q_\beta q_\gamma\omega_\beta\omega_\gamma}{2}\int_{\mathcal{C}}\frac{dz}{2\pi i\sqrt{z}}\frac{1}{(z-\omega_\beta^2)(z-\omega_\gamma^2)}\frac{1}{1-\eta\left(q^T\frac{\Omega}{z-\Omega^2}q\right)}\\[5pt]
    &=\frac{\delta_{\beta\gamma}}{2}+\frac{\eta\,q_\beta q_\gamma\omega_\beta\omega_\gamma}{2(\omega_\beta^2-\omega_\gamma^2)}\int_{\mathcal{C}}\frac{dz}{2\pi i\sqrt{z}}\left[\frac{1}{z-\omega_\beta^2}-\frac{1}{z-\omega_\gamma^2}\right]\frac{1}{1-\eta\left(q^T\frac{\Omega}{z-\Omega^2}q\right)}\\[5pt]
    &=\frac{\delta_{\beta\gamma}}{2}+\frac{\eta\,q_\beta q_\gamma}{2(\omega_\beta^2-\omega_\gamma^2)}\Big[\omega_\gamma l(\omega_\beta)-\omega_\beta(\omega_\gamma)\Big],
\end{split}\end{align}
where $l(\omega)$ is defined in Eq.~(\ref{eqn:appB_l1}) with $D\cos\theta\to \eta$. Replacing $q_\beta$ using Eq.~(\ref{eqn:AppB_qvector}) and with $\theta=0$, we arrive at
\begin{equation}
    \braket{\hat{p}_\beta\hat{p}_\gamma}=\frac{\delta_{\beta\gamma}}{2}+\frac{D \,g_\beta g_\gamma}{2(\omega_\beta^2-\omega_\gamma^2)\big[\omega_\beta+Dl(\omega_\beta)\big]\big[\omega_\gamma+D l(\omega_\gamma)\big]}\Big[\omega_\gamma l(\omega_\beta)-\omega_\beta(\omega_\gamma)\Big]
\end{equation}
Defining $y=\omega/D$ we then obtain
\begin{equation}
     \braket{\hat{p}_\beta\hat{p}_\gamma}=\frac{\delta_{\beta\gamma}}{2}+\frac{\,g_\beta g_\gamma/D^2}{2(y_\beta^2-y_\gamma^2)\big[y_\beta+m(y_\beta)\big]\big[y_\gamma+m(y_\gamma)\big]}\Big[y_\gamma m(y_\beta)-y_\beta m(y_\gamma)\Big]
\end{equation}
Similar manipulations lead to 
\begin{align}\begin{split}
    \braket{\hat{x}_\beta\hat{x}_\gamma}&=\frac{\delta_{\beta\gamma}}{2}+\frac{\eta \,q_\beta q_\gamma}{2}\int_{\mathcal{C}}\frac{\sqrt{z}dz}{2\pi i}\frac{1}{(z-\omega_\beta^2)(z-\omega_\gamma^2)}\frac{1}{1-\eta\left(q^T\frac{\Omega}{z-\Omega^2}q\right)}\\[5pt]
    &=\frac{\delta_{\beta\gamma}}{2}+\frac{\eta \,q_\beta q_\gamma}{2(\omega_\beta^2-\omega_\gamma^2)}\int_{\mathcal{C}}\frac{dz}{2\pi i\sqrt{z}}\left[\frac{\omega_\beta^2}{z-\omega_\beta^2}-\frac{\omega_\gamma^2}{z-\omega_\gamma^2}\right]\frac{1}{1-\eta\left(q^T\frac{\Omega}{z-\Omega^2}q\right)}\\[5pt]
   &=\frac{\delta_{\beta\gamma}}{2}+\frac{\eta \,q_\beta q_\gamma}{2(\omega_\beta^2-\omega_\gamma^2)}\Big[\omega_\beta l(\omega_\beta)-\omega_\gamma l(\omega_\gamma)\Big]\\[5pt]
    \rightarrow\braket{\hat{x}_\beta\hat{x}_\gamma}&=\frac{\delta_{\beta\gamma}}{2}+\frac{g_\beta g_\gamma/D^2}{2(y_\beta^2-y_\gamma^2)\big[y_\beta+m(y_\beta)\big]\big[y_\gamma+m(y_\gamma)\big]}\Big[y_\beta m(y_\beta)-y_\gamma m(y_\gamma)\Big]
\end{split}\end{align}

\section{NUMERICAL SOLUTION OF INTEGRAL EQUATION}\label{appC}
Here we outline the numerical solution of the integral equation
\begin{equation}
    m(y)=\int_0^\infty\frac{dz}{\pi}\frac{y}{y^2+z^2}\frac{1}{1+2b\int_0^{y_c}\frac{x^{s+1}dx}{(x^2+z^2)\big[x+m(x)\big]^2}}.
\end{equation}
We first solve for this equation at fixed $y_c=\omega_c\cos\theta/D=\omega_c/\Delta\times(\Delta\cos\theta/D)$ and then systematically improve its value. We now define the sequence of functions $m_n(x)$ according to
\begin{equation}
    m_{n+1}(y)=\int_0^\infty\frac{2dz}{\pi}\frac{y}{y^2+z^2}\frac{1}{1+2b\int_0^{\infty}\frac{x^{s+1}dx}{(x^2+z^2)\big[x+m_{n}(x)\big]^2}}.
\end{equation}
If we choose $m_0^I(y)=1$ then $m_1^I(y)<m_0^I(y)$. Furthermore, we also have that if $m_{n+1}(y)\leq m_n(y)$ then $m_{n+2}(y)\leq m_{n+1}(y)$ pointwise. Thus, with our choice of $m_0^I(y)$ each progressive iteration in the sequence is pointwise decreasing. Conversely, if we choose $m_0^{II}(y)=0$ then $m_1^{II}(y)>m_0^{II}(y)$ and if $m_{n+1}(y)\geq m_n(y)$ then $m_{n+2}(y)\geq m_{n+1}(y)$. Thus the sequence $m_n^{II}(y)$ is pointwise increasing. Numerically we find that both sequences converge to the same function. Away from the critical point, between $3$ and $10$ iterations are enough to get convergence with a pointwise relative error of $10^{-6}$ between iterations. Closer to the critical point the number of iterations grows. 

For the outer integral we have 
\begin{align}\begin{split}
    \int_0^{\infty}\frac{2ydz}{\pi(z^2+y^2)}&\frac{1}{1+2b\int_0^{y_c}\frac{x^{s+1}dx}{(x^2+z^2)\big[x+m(x)\big]^2}}\\[10pt]
    &=\int_{0}^{1000}+\int_{1000}^{\infty}\\[10pt]
    &\approx\int_{0}^{1000}\frac{2ydz}{\pi(z^2+y^2)}\frac{1}{1+2b\int_0^{\infty}\frac{x^{s+1}dx}{(x^2+z^2)\big[x+m(x)\big]^2}}+\int_{1000}^{\infty}\frac{2ydz}{\pi(z^2+y^2)}\\[10pt]
    &=\int_{0}^{1000}\frac{2ydz}{\pi(z^2+y^2)}\frac{1}{1+2b\int_0^{y_c}\frac{x^{s+1}dx}{(x^2+z^2)\big[x+m(x)\big]^2}}+\frac{2}{\pi}\left[\frac{\pi}{2}-\arctan\left(\frac{1000}{y}\right)\right]
\end{split}\end{align}
and so the correction can be sizable when $y\approx 1000$. Thus, we set up our numerical iteration scheme as follows:
\begin{equation}
    m_{n+1}(y)=\int_0^{1000}\frac{2dz}{\pi}\frac{y}{y^2+z^2}\frac{1}{1+2b\left[\int_0^{y_c}\frac{x^{s+1}dx}{(x^2+z^2)\big[x+m_{n}(x)\big]}\right]}+\frac{2}{\pi}\left[\frac{\pi}{2}-\arctan\left(\frac{1000}{y}\right)\right].
\end{equation}
By varying the upper limits between $100$ and $1000$ we empirically found no appreciable changes in the converged $m(y)$. To perform the numerical integrations, we first changed variables to $z=\tilde{z}^{p}$ and $y=\tilde{y}^p$, with $p=2.5$ and then used Gauss-Legendre quadratures with $8000$ nodes in the variables $\tilde{y},\tilde{z}$. This transformation allows for a finer resolution of the inner and outer integrands in the regions of small $y,z$, which are crucial for a quantitatively accurate determination of the low energy properties of $m(y)$ near criticality. In a few cases ($s=0.8$ and close to the corresponding critical point), we had to increase $p\to 3.5$ and the number of Gauss-Legendre nodes to get convergence in the small $y$ region.

We begin this process at fixed $y_c^{(0)}=\omega_c/\Delta=10$. Once we obtain a converged $m(y)$ we calculate $\Delta\cos\theta/D$ and improve our estimate of $y_c$. About $5$ iterations are enough to converge $y_c$ to within a relative error of $0.5\%$. To avoid doing this for every single value of $\alpha_\Delta$, we sweep increasing $\alpha_\Delta$, using the previous converged values of $y_c,m(y)$ as inputs for a subsequent $\alpha_\Delta$.

\begin{figure}
    \centering
    \includegraphics[width=0.95\linewidth]{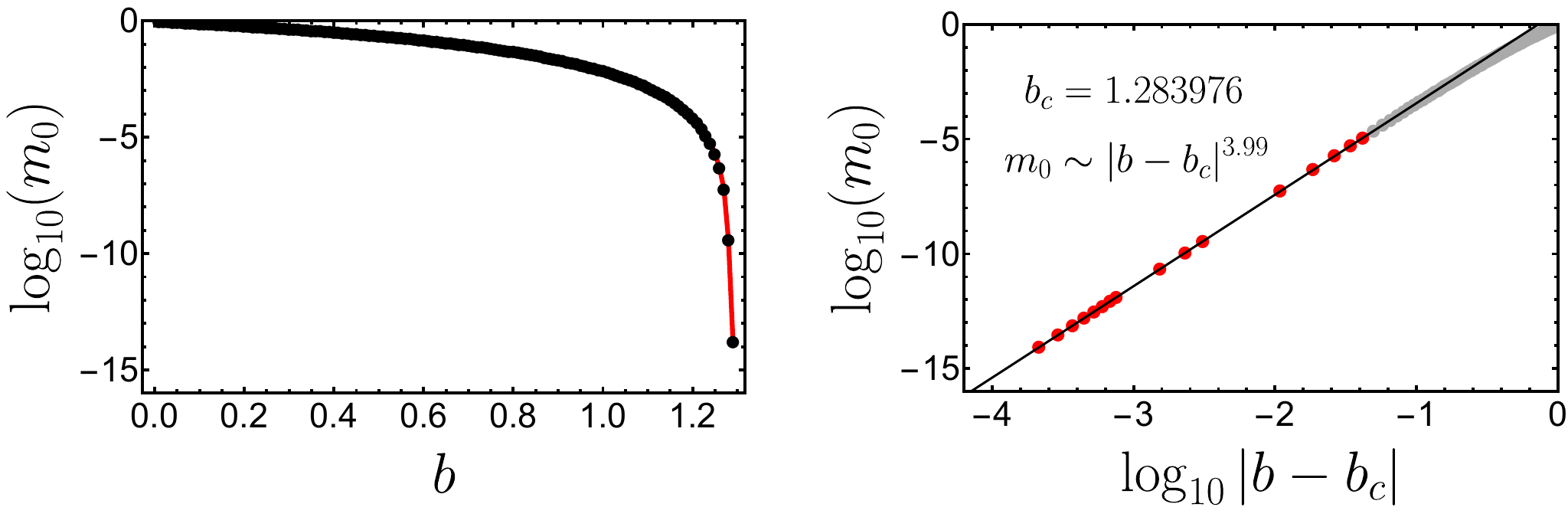}
    \caption{(Left) Zero frequency value $m_0$ as a function of $b$, indicating the approach to the critical $b_c$ for $s=0.8$. (Right) $m_0$ vs $|b-b_c|$ in a logarithmic plot after the optimal $b_c$ has been determined, for $s=0.8$.}
    \label{fig:SMCritExponent}
\end{figure}

As $b$ is sweeped in finite increments, there is a  pair of values $b_N,b_{N+1}$ (for some integer $N$) such the zero frequency value $m_0^N=m(0)$ jumps from a small value to $m_0^{N+1}=0$ within machine precision, indicating that the critical point $b_c$ is located in the interval $I=[b_N,b_{N+1}]$ [see Fig.~\ref{fig:SMCritExponent}(left) for $s=0.8$]. To extract the dependence between $m_0$ and $|b-b_c|$, we perform a sweep within $I$ with increased precision to constrain further the location of $b_c$ and repeat this procedure a few (2 to 4, depending on $s$) times. As we get closer to $b_c$, we verify convergence of $m_0$ by increasing the number of Legendre-Gauss nodes in the numerical integrals and increasing $p$. We optimize over the value of $b$ which leads to the best linear fit of $\log(m_0)$ vs $\log|b-b_c|$ and extract the associated exponent, as depicted in Fig.~\ref{fig:SMCritExponent}(right) for $s=0.8$.

Using this procedure, we find numerically for $s-0.2,0.3,0.4,0.5,0.6,0.7,0.8$ the exponents $0.986\pm0.003,1.0036\pm0.0002 ,1.0200\pm 0.0008,1.0913\pm0.0005,1.5134\pm 0.0007,2.3325\pm 0.0002,3.986\pm 0.001$, in good agreement with the theoretical calculation $m_0\sim|b-b_c|^{\text{max}\big(1,\frac{s}{1-s}\big)}$.

\section{CRITICAL COUPLING COMPARISONS}\label{appD}

\begin{figure}
    \centering
    \includegraphics[width=0.98\linewidth]{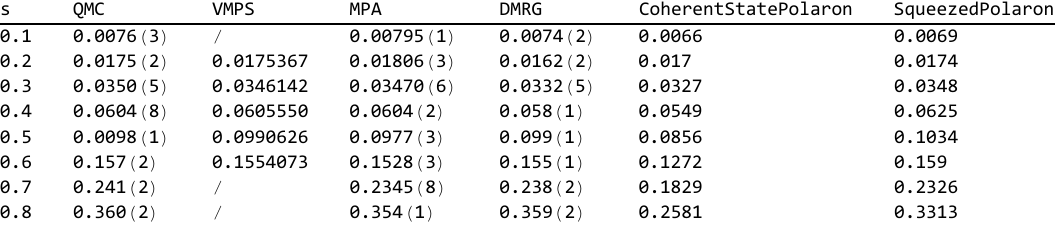}
    \caption{Critical values of $\alpha$ determined via various methods, all at $\omega_c/\Delta=0.1$.}
    \label{fig:SM_critvalues}
\end{figure}

In this section we compare the critical values of the coupling constant predicted by the squeezed polaron against the values obtained via other numerical or variational methods. The results are 
summarized in Fig.~\ref{fig:SM_critvalues}. In contrast to the main body of the paper, here we quote the values of $\alpha$ instead of $\alpha_\Delta=\alpha(\omega_c/\Delta)^{1-s}$, as is typically done in the literature. The numerical methods explored include Quantum Monte Carlo (QMC)\cite{Winter2009}, variational matrix-product states (VMPS)~\cite{2012Guo}, variational multiple polaron ansatz (VPA)~\cite{SHEN20235}, and DMRG~\cite{Wong2008}, all of which are obtained at the finite value $\omega_c=10\Delta$. Furthermore, for simplicity on the analytical side, the values we quote are obtained by first calculating the values $\alpha_\Delta$ in the scaling limit where $\omega_c/\Delta\to \infty$ and then multiplying them by $(0.1)^{1-s}$, instead of solving all the self-consistency conditions at $\omega_c/\Delta-10$. 

\section{LOW FREQUENCY BEHAVIOUR OF M(Y) AT THE CRITICAL POINT}\label{appE}
In this section we begin from the integral equation for $m(y)$
\begin{equation}
    m(y)=\int_0^{\infty}\frac{2}{\pi}\frac{y\,dz}{z^2+y^2}\frac{1}{1+2b\int_0^{y_c}\frac{x^{s+1}dx}{(x^2+z^2)\big[x+m(x)\big]^2}}
\end{equation}
and extract the low frequency behaviour at the critical point. We assume that at small $y$, $m(y)\approx B y^p$ follows a power law. Since we are interested in small $y$ it follows that the region of small $z$ is the most relevant for the outer integral. This in turn means that the region of small $x$ is the most relevant for the inner integral. To proceed, we separate the integral region
\begin{equation}
    \int_0^{y_c}\frac{x^{s+1}dx}{(x^2+z^2)\big[x+m(x)\big]^2}= \int_0^{\Lambda}\frac{x^{s+1}dx}{(x^2+z^2)\big[x+m(x)\big]^2}+ \int_{\Lambda}^{y_c}\frac{x^{s+1}dx}{(x^2+z^2)\big[x+m(x)\big]^2},
\end{equation}
where $z\ll\Lambda\ll 1$ is chosen so that in the region $x\leq \Lambda$ the function $m(x)$ attains its low frequency form $m(x)\approx Bx^p$. We further assume that $p<1$ so that $x^p$ dominates over $x$ in the left integral. Since $\Lambda\gg z$, we can neglect $z$ in the right integral. We thus have 
\begin{align}\begin{split}
     \int_0^{y_c}\frac{x^{s+1}dx}{(x^2+z^2)\big[x+m(x)\big]^2}&\approx\int_0^{\Lambda}\frac{x^{s+1-2p}dx}{(x^2+z^2)B^2}+ \int_{\Lambda}^{y_c}\frac{x^{s-1}dx}{\big[x+m(x)\big]^2}\\[5pt]
     &=\frac{1}{B^2 z^{2p-s}}\int_0^{\Lambda/z}\frac{x^{s+1-2p}dx}{(x^2+1)}+ \int_{\Lambda}^{y_c}\frac{x^{s-1}dx}{\big[x+m(x)\big]^2}.
\end{split}\end{align}
If we assume $2p>s$ then the left integral is convergent as $\Lambda/z\to \infty$, so that we can re-express the full integral as
\begin{align}\begin{split}
     \int_0^{y_c}\frac{x^{s+1}dx}{(x^2+z^2)\big[x+m(x)\big]^2}&\approx\frac{1}{B^2 z^{2p-s}}\left(\int_0^{\infty}\frac{x^{s+1-2p}dx}{(x^2+1)}-\int_{\Lambda/z}^{\infty}\frac{x^{s+1-2p}dx}{(x^2+1)}\right)+ \int_{\Lambda}^{y_c}\frac{x^{s-1}dx}{\big[x+m(x)\big]^2}\\[5pt]
     &\approx\frac{1}{B^2 z^{2p-s}}\left(\int_0^{\infty}\frac{x^{s+1-2p}dx}{(x^2+1)}+\frac{z^{2p-s}}{(s-2p)\Lambda^{2p-s}}\right)+ \int_{\Lambda}^{y_c}\frac{x^{s-1}dx}{\big[x+m(x)\big]^2}\\[5pt]
     &=\frac{1}{B^2 z^{2p-s}}\int_0^{\infty}\frac{x^{s+1-2p}dx}{(x^2+1)}+\left( \int_{\Lambda}^{y_c}\frac{x^{s-1}dx}{\big[x+m(x)\big]^2}+\frac{1}{B^2(2p-s)\Lambda^{2p-s}}\right).
\end{split}\end{align}
In the term in parentheses we can take the limit $\Lambda\to 0$ safely, which contributes a constant. Since $2p>s$, the first term dominates at small $z$. Thus, the outer integral becomes
\begin{equation}
    m(y)\approx \frac{B^2}{2b \int_0^{\infty}\frac{x^{s+1-2p}dx}{(x^2+1)}}\int_0^\infty\frac{2}{\pi}\frac{y\,dz}{z^2+y^2}\times z^{2p-s}=y^{2p-s}\frac{B^2}{b \int_0^{\infty}\frac{x^{s+1-2p}dx}{(x^2+1)}}\int_0^\infty\frac{1}{\pi}\frac{z^{2p-s}dz}{z^2+1}.
\end{equation}
From self-consistency with our assumption that $m(y)\approx By^p$ we thus have
\begin{align}
    \begin{split}
        p=2p-s\to p&=s\\[5pt]
        B=\frac{B^2}{b \int_0^{\infty}\frac{x^{s+1-2p}dx}{(x^2+1)}}\int_0^\infty\frac{1}{\pi}\frac{z^{2p-s}dz}{z^2+1}\to B&=\pi b\frac{\int_0^{\infty}\frac{x^{1-s}dx}{(x^2+1)}}{\int_0^\infty\frac{z^{s}dz}{z^2+1}}=\pi b \cot(\pi s/2)
    \end{split}
\end{align}
This solution for $p$ is also consistent with our assumption $2p>s$. In summary we have $\boxed{m(y)\approx\pi b \cot(\pi s/2)y^s}$.

\section{SCALING BEHAVIOUR OF M(Y) AT LOW FREQUENCY}\label{appF}
\begin{figure}
    \centering
    \includegraphics[width=0.49\linewidth]{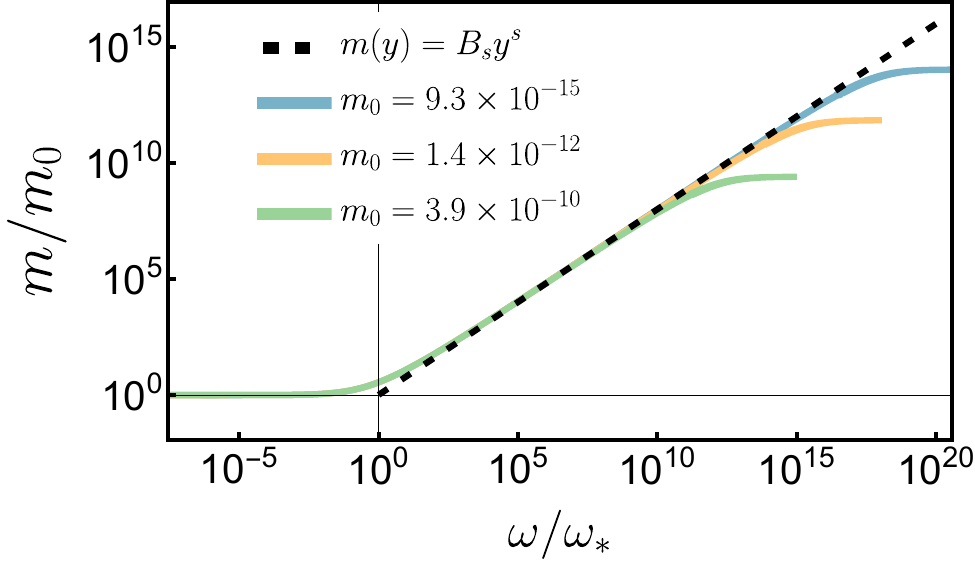}
    \includegraphics[width=0.49\linewidth]{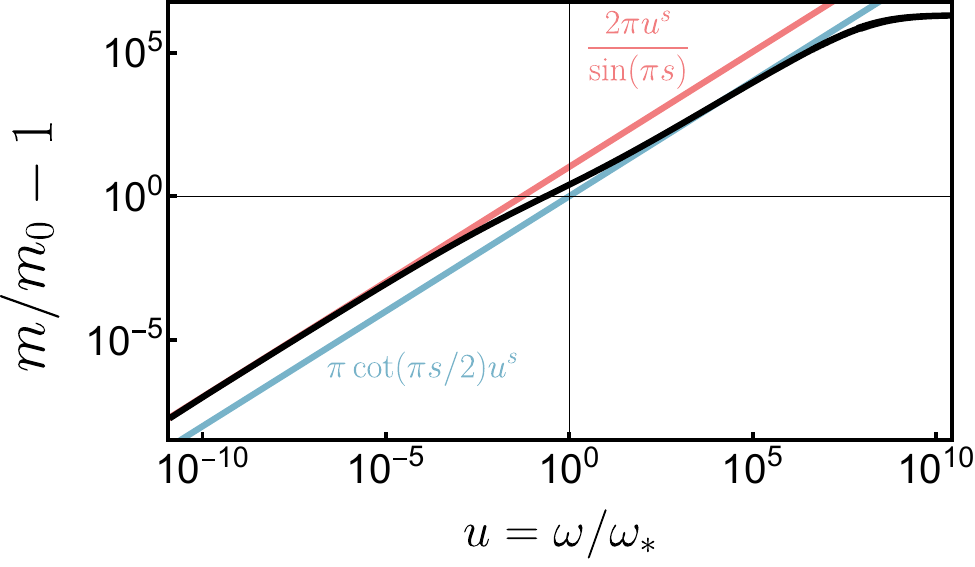}
    \caption{(Left) Numerically determined $m(y)$ for $s=0.8$ and three different values of $m_0=9\times10^{-15},1\times 10^{-12},4\times 10^{-10}$ (corresponding to $b=1.2837,1.283,1.28$, respectively). The curves collapse in the small frequency region when scaled appropriately. Dashed line is the analytical result at criticality $m(y)=\pi b \cot(\pi s/2)y^s$ (Right) Asymptotic behaviours of $r(u)-1=m/m_0-1$ compared with the numerical solution of $m(y)$ for $s=0.8$ and $m_0=5\times 10^{-7}$ ($b=1.27$).}
    \label{fig:SM_scalingfunction}
\end{figure}

In this section we begin from the integral equation for $m(y)$
\begin{equation}
    m(y)=\int_0^{\infty}\frac{2}{\pi}\frac{y\,dz}{z^2+y^2}\frac{1}{1+2b\int_0^{y_c}\frac{x^{s+1}dx}{(x^2+z^2)\big[x+m(x)\big]^2}}
\end{equation}
and reduce it to the scaling function when $b\approx b_c$ and $y\ll 1$. Since we are interested in the regime $y\ll 1$, this means that most of the contribution to the outer integral comes from $z\ll1$. This implies that most of the contribution to the inner integral comes mostly from the region $x\ll 1$. In this region, we expect that $m(x)\gg x$. To perform this more quantitatively, we write the inner integral as
\begin{equation}
    \int_0^{y_c}\frac{x^{s+1}dx}{(x^2+z^2)\big[x+m(x)\big]^2}=\int_0^{\Lambda}\frac{x^{s+1}dx}{(x^2+z^2)\big[x+m(x)\big]^2}+\int_\Lambda^{y_c}\frac{x^{s+1}dx}{(x^2+z^2)\big[x+m(x)\big]^2}
\end{equation}
with $z\ll\Lambda\ll 1$. In the first integral we replace $x+m(x)\approx m(x)$ since $m(x)$ dominates the low frequency region. In the right integral, we instead can set $z=0$ so we have
\begin{equation}
    \int_0^{y_c}\frac{x^{s+1}dx}{(x^2+z^2)\big[x+m(x)\big]^2}=\int_0^{\Lambda}\frac{x^{s+1}dx}{(x^2+z^2)\big[m(x)\big]^2}+\int_\Lambda^{y_c}\frac{x^{s-1}dx}{\big[x+m(x)\big]^2}
\end{equation}
As in the determination of the critical behaviour of $m(y)$, the left integral converges for large $x$ and dominates the behaviour at small $z$ so we extend the integration limit to $\infty$ and omit the right integral, since it is just a $z$ independent constant. Thus, in the end we have
\begin{equation}
    m(y)\approx \int_0^{\infty}\frac{2}{\pi}\frac{y dz}{y^2+z^2}\frac{1}{2b\int_0^{\infty}\frac{x^{s+1}\,dx}{(x^2+z^2)\big[m(x)\big]^2}}.
\end{equation}
We express $m(y)=m_0 r(y/y_*)$, so that $r(0)=1$. Plugging in this form and rescaling the $z,x$ integrals appropriately, we end with
\begin{equation}
    r(u)=\int_0^{\infty}\frac{2}{\pi}\frac{u\,dz}{u^2+z^2}\frac{1}{\frac{2by_*^s}{m_0}\int_0^{\infty}\frac{x^{s+1}\,dx}{(x^2+z^2)\big[r(x)\big]^2}}.
\end{equation}
If we choose $by_*^s/m_0=1$ we end up with the following integral equation for $r(u)$
\begin{equation}
    \boxed{r(u)=\int_0^{\infty}\frac{2}{\pi}\frac{u\,dz}{u^2+z^2}\frac{1}{2\int_0^{\infty}\frac{x^{s+1}\,dx}{(x^2+z^2)\big[r(x)\big]^2}}}.
\end{equation}
with the condition $\boxed{r(0)=1}$. These two last equations are independent of parameters of the model. Then, the actual $m(y)$ satisfies
\begin{equation}
    m(y)=m_0\,r\left(\frac{m_0^{1/s}y}{b^{1/s}}\right).
\end{equation}
Remembering that $y=\omega/D$, we obtain the scale $\omega_*=Dm_0^{1/s}/b^{1/s}$. Figure~\ref{fig:SM_scalingfunction} (left panel) shows that $m(y)$ indeed displays scaling behaviour as $m_0\to 0$ by showing the numerically determined $m(y)$ for $s=0.8$ and different values of $m_0$ (i.e. different $b$) and demonstrating that they collapse when the variables are scaled appropriately. From the perspective of the scaling function $r(u)$, the critical region is at $u\gg 1$ and in fact the integral equation for $r(u)$ does asymptote to $r(u)\to\pi \cot(\pi s/2)u^s$ as $u\gg 1$. The behaviour at $u\ll 1$ is accessed at the lowest frequencies ($\omega\ll\omega_*$) and corresponds to probing the $m(y)\approx m_0$ regime. From the integral equation for $r(u)$ we get that $r(u)\approx 1+\frac{2\pi}{\sin(\pi s)}u^s$ when $u\ll 1$. Both analytically derived asymptotic behaviours are verified by comparison to the numerical solution of $m(y)$ for $s=0.8$ and $m_0=5\times10^{-7}$ ($b=1.27$) [see Fig.~\ref{fig:SM_scalingfunction} (right panel)].  

\section{CORRELATION LENGTH EXPONENT}\label{appG}
The critical exponent $\nu$ can be extracted from the equation for the scaling function $r(u)$, which interpolates between the critical regime and the non-critical regime. The main idea comes from noticing that the equation for $r(u)$
\begin{equation}
    r(u)=\int_0^{\infty}\frac{2}{2\pi}\frac{u\,dz}{(u^2+z^2)\int_0^{\infty}\frac{x^{s+1}dx}{(x^2+z^2)\big[r(x)\big]^2}}
\end{equation}
has as exact solution $r_c(u)=\pi \cot(\pi s/2) u^s$ if we omit the requirement $r(0)=1$, and this corresponds to being at criticality. The boundary condition $r(0)=1$ then means that at very low energies the system enters the delocalized phase, even if at $u\gg 1$ it displays critical behaviour [$r(u)\to r_c(u)$]. This further means that at large $u$, $r(u)$ is slightly perturbed from $r_c(u)$, and we assume that this perturbation is a power law
\begin{equation}\label{app:CritExp:Ansatz}
    r(u)=Bu^s\left(1+\frac{C}{u^t}\right),
\end{equation}
where $B=\pi \cot(\pi s/2)$, $C$ is a proportionality constant and $t>0$ is an exponent that we seek to determine. Note that $t>0$ because we want a small perturbation about $r_c(u)$. Since at $u\gg 1$ the outer and inner integrals are dominated by $z\gg1$ and $x\gg1$, respectively, we simplify the inner integral as follows
\begin{equation}
    \int_0^{\infty}\frac{x^{s+1}dx}{(x^2+z^2)\big[r(x)]^2}\approx \int_0^{\infty}\frac{x^{1-s}dx}{B^2(x^2+z^2)(1+\frac{2C}{x^{t}})}\approx \frac{1}{B^2}\left(\int_0^{\infty}\frac{x^{1-s}dx}{x^2+z^2}-2C\int_0^{\infty}\frac{x^{1-s-t}dx}{x^2+z^2}\right),
\end{equation}
where we are assuming that the region near $x\sim 1$ only contributes subleadingly (in $z$) to the integral. Plugging this into the outer integral leads to
\begin{align}\begin{split}
    r(u)&\approx\int_0^{\infty}\frac{u dz}{\pi(z^2+u^2)}\times\frac{1}{\frac{1}{B^2}\left(\int_0^{\infty}\frac{x^{1-s}dx}{x^2+z^2}-2C\int_0^{\infty}\frac{x^{1-s-t}dx}{x^2+z^2}\right)}\\[10pt]
    &\approx \int_0^{\infty}\frac{u dz}{\pi(z^2+u^2)}\times\frac{1}{\frac{1}{B^2}\int_0^{\infty}\frac{x^{1-s}dx}{x^2+z^2}}\times\left(1+2C\frac{\int_0^{\infty}\frac{x^{1-s-t}dx}{x^2+z^2}}{\int_0^{\infty}\frac{x^{1-s}dx}{x^2+z^2}}\right)\\[10pt]
    &=B^2 u^s\frac{\int_0^{\infty}\frac{z^{s}dz}{z^2+1}}{\int_0^{\infty}\frac{x^{1-s}dx}{x^2+1}}\left(1+\frac{2C}{u^t}\frac{\int_0^{\infty}\frac{x^{1-s-t}dx}{x^2+1}\times\int_0^{\infty}\frac{z^{s-t}dz}{z^2+1}}{\int_0^{\infty}\frac{x^{1-s}dx}{x^2+1}\times\int_0^{\infty}\frac{z^{s}dz}{z^2+1}}\right).
\end{split}\end{align}
Consistency with the original assumption leads to $B=\pi \cos(\pi s/2)$ as before and
\begin{align}\begin{split}
    2\frac{\int_0^{\infty}\frac{x^{1-s-t}dx}{x^2+1}\times\int_0^{\infty}\frac{z^{s-t}dz}{z^2+1}}{\int_0^{\infty}\frac{x^{1-s}dx}{x^2+1}\times\int_0^{\infty}\frac{z^{s}dz}{z^2+1}}=1&\longrightarrow\sin(\pi s)=\cos\left[\frac{\pi}{2}(s-t)\right]\sin\left[\frac{\pi}{2}(s+t)\right]\\[5pt]
    &\longrightarrow\boxed{\sin(\pi s)=\sin(\pi t)}
\end{split}\end{align}
The constant $C$ is undetermined by these arguments. The equation for $t$ has many solutions but the important one is the one with the lowest value of $t>0$, which leads to
\begin{equation}
    \boxed{t=\text{min}(s,1-s)}
\end{equation}
To relate this to the distance of the control parameter from its critical value, $b-b_c$, we use the asymptotic form for $r(u)$ to obtain
\begin{align}\begin{split}
    m(y)&=m_0r(y/y_*)\approx \pi\cot(\pi s/2) m_0\left(y/y_*\right)^s\left[1+\frac{C}{(y/y_*)^t}\right]\\[5pt]
    m(y)&\approx \pi b\cot(\pi s/2)y^s\left(1+\frac{m_0^{t/s}C}{b^{t/s}y^t}\right).
\end{split}\end{align}
The leading behaviour is independent of $m_0$, which is expected since $m_0=0 $ at criticality. The subleading behaviour thus quantifies the deviation from criticality, and it manifestly vanishes when $m_0$ goes to $0$. This behaviour for $m(y)$, which is attained at frequencies $\omega>\omega_*$, occurs at relatively high energies (although still $y\ll 1$), in which case it is safe to consider perturbation theory about the critical point
\begin{equation}
    m(y)=m_c(y)+|b-b_c|m_c^{(1)}(y)+...
\end{equation}
Comparing the dependence of the subleading terms we thus obtain that
\begin{equation}
    m_0^{t/s}\propto |b-b_c|\to m_0\propto |b-b_c|^{s/t}\propto|b-b_c|^{\text{max}\left(1,\frac{s}{1-s}\right)},
\end{equation}
This is Eq.~(\ref{eqn:CriticalProperties:CritExp}). Using Eq.~(\ref{eqn:SqueezedPolaron:CouplingConstant}) in the delocalized phase ($\theta=0$), we have that $|\alpha_\Delta-\alpha_\Delta^c|\propto |b-b_c|$ (the integrals in the exponentials are convergent, so they do not modify this relation). Thus
\begin{equation}
    \omega^*\sim m_0^{1/s}\sim |b-b_c|^{1/t}\sim|\alpha_\Delta-\alpha_\Delta^c|^{1/t}\longrightarrow \boxed{\nu^{-1}=t=\min(s,1-s).}
\end{equation} 
At $s=0.5$ both these exponents are degenerate, so we expect instead logarithmic corrections.

\section{EFFECTIVE HAMILTONIAN FOR SCATTERING}\label{appH}

The Hamiltonian, after application of the correlated displacement unitary, is given by Eq.~(\ref{eqn:appendix:CorrelatedHamiltonian}), which we copy here for reference purposes. 
\begin{align}\begin{split}
    \hat{H}_c=\mathcal{D}_c^\dagger\hat{H}\mathcal{D}_c&=\frac{\Delta}{2}\left[\hat{\sigma}_+e^{-\sum_\beta q_\beta(\hat{a}_\beta-\hat{a}_\beta^\dagger)}+\hat{\sigma}_- e^{\sum_\beta q_\beta(\hat{a}_\beta-\hat{a}_\beta^\dagger)}\right]+\sum_\beta(g_\beta-\omega_\beta q_\beta)\hat{s}_z(\hat{a}_\beta+\hat{a}_\beta^\dagger)+\sum_\beta \omega_\beta\hat{a}^\dagger_\beta\hat{a}_\beta\\
    &+\sum_\beta \frac{(g_\beta-\omega_\beta q_\beta)^2}{4\omega_\beta}-\sum_\beta\frac{g_\beta^2}{4\omega_\beta}
\end{split}\end{align}
We now assume that we are in the delocalized phase and that the minimization procedure has already been performed, so that $q_\beta$ are given by
\begin{equation}
    q_\beta=\frac{g_\beta}{\omega_\beta+Dm(\omega_\beta/D)}
\end{equation}
with no spin rotation nor uncorrelated displacements. For notational simplicity we will now write $m(\omega_\beta/D)=m_\beta$. If we normal-order the exponentials with respect to the variational gaussian state $\ket{\phi}$ and expand them to second order, then the resulting Hamiltonian is 
\begin{align}\begin{split}
    \hat{H}_c&\approx D\left(\hat{s}_x+\frac{1}{2}\right)+:\sum_\beta\hat{a}_\beta^\dagger\hat{a}_\beta-D\hat{s}_x\sum_{\beta\beta'}q_\beta q_{\beta'} \hat{p}_\beta\hat{p}_{\beta'}:+E_{\text{gnd}}\\
    &-iD\left\{\underbrace{\left[\sum_\beta\left(\frac{m_\beta\hat{x}_\beta+i\hat{p}_\beta}{\sqrt{2}}\right)q_\beta\right]}_{\hat{B}}\hat{\sigma}_+^x -\underbrace{\left[\sum_\beta\left(\frac{m_\beta\hat{x}_\beta-i\hat{p}_\beta}{\sqrt{2}}\right)q_\beta\right]}_{\hat{B}^\dagger}\hat{\sigma}_-^x\right\},
\end{split}\end{align}
where $\hat{\sigma}_{\pm}^x=\hat{s}_y\pm i\hat{s}_z$ are raising/lowering operators for $\hat{s}_x$ and $E_{\text{gnd}}$ is the variational ground state energy. Furthermore, the operator $\hat{B}$ that accompanies $\hat{\sigma}_+^x$ annihilates the variational ground state $\ket{\phi}$. Thus $\ket{-x}\ket{\phi}$ is the exact ground state of this Hamiltonian. We can then ask how is sincle excitation scattered by the impurity spin. When the spin is flipped, the $\hat{p}_k\hat{p}_{k'}$ term changes sign, so that it will create particls out of $\ket{-x}\ket{\phi}$. We project onto the particle conserving sector to avoid this process. Because under this approximation there is no multiparticle production, the $1\to 1$ channel should be unitary on its own and the end result is just that the scattered photon gets phase shifted. This is what we are after. We can obtainthis phase shift via the formalism outlined in~\cite{Zarand2003,Borda2007} for magnetic impurities, which requires the time-ordered Green's function
\begin{equation}
    G^F_{kk'}(E)=\braket{\hat{a}_k\frac{1}{E-(\hat{H}_c-E_{\text{gnd}})+i\epsilon} \hat{a}^\dagger_{k'}}-\braket{\hat{a}_{k'}^\dagger\frac{1}{E+\hat{H}_c-E_{\text{gnd}}-i\epsilon} \hat{a}_{k}},
\end{equation}
where the bracket is taken with respect to the variational ground state $\ket{-x}\ket{\phi}$. The $T$ matrix elements are then given by
\begin{equation}
    T(E,E')=-2\pi\delta(E-E')\Big[(E-\omega_k)(E-\omega_{k'})G^F_{kk'}(E)\Big]_{\omega_k,\omega_{k'}\to E}
\end{equation}
When we amputate the incoming/outgoing photon (the $E-\omega_k$ factors) and put the amplitude on-shell (at positive energy $E=\omega_k$) the second term of $G^F(E)$ will not contribute , so we will focus only on
\begin{equation}
    G_{kk'}(E)=\braket{\hat{a}_k\frac{1}{E-(\hat{H}_c-E_{\text{gnd}})+i\epsilon} \hat{a}^\dagger_{k'}}.
\end{equation}
Since only the single excitation sector is coupled, we obtain
\begin{equation}
    G_{kk'}(E)=\braket{\hat{a}_k\hat{G}_{-}\hat{a}^\dagger_{k'}}_\phi+D^2\frac{\braket{\hat{a}_k\hat{G}_-\hat{B}^\dagger}_\phi\times\braket{\hat{B}\hat{G}_-a^\dagger_{k'}}_\phi}{E-D-D^2\braket{\hat{B}\hat{G}_-\hat{B}^\dagger}_\phi},
\end{equation}
where the averages are now against only the bosonic state $\ket{\phi}$ and $\hat{G}_{-}=(E-\hat{H}_b+i\epsilon)$ and $\hat{H}_b=:\sum_{k}\omega_k\hat{a}_k^\dagger\hat{a}_k+D\sum_k\hat{p}_k\hat{p}_{k'}q_k q_{k'}/2:$ is normal ordered with respect to $\ket{\phi}$.  Since $\hat{H}_b$ is quadratic in boson operators, the many-body Green's function $G_{kk'}(E)$ can be expressed entirely in terms of single-particle quantities. To do this, we introduce the Nambu vector $\hat{v}^T=(\hat{x}_k\hspace{0.2cm}\hat{p}_k)$, where the $k$'s act as indices. Then
\begin{equation}
    \tilde{G}(E)=\braket{\phi|\hat{v}\,G_-(E)\hat{v}^T|\phi}=\frac{1}{2}\underbrace{\begin{pmatrix}
        W^{-1/2}&0\\0&W^{1/2}
    \end{pmatrix}}_{f(W)}
    \frac{1}{E-\Omega+i\epsilon}\begin{pmatrix}
        \Omega&i\\-i& \Omega^{-1}
    \end{pmatrix}
    \begin{pmatrix}
            W^{-1/2}&0\\0&W^{1/2},
        \end{pmatrix}
\end{equation}
where $W=\text{diag}(\omega_k)$ is a diagonal matrix of frequencies, $\Omega^2=W^2+DW^{1/2}qq^TW^{1/2}$ are the frequencies of the normal modes of $\hat{H}_b$ and we have introduced $f(W)$ for simplicity . The various two-point functions appearing in $G_{kk'}(E)$ are then obtained via appropriate projections. Using resolvent methods, we can express this as
\begin{align}\begin{split}
    \tilde{G}(E)&=\frac{f(W)}{2}\oint_{\mathcal{C}}\frac{du}{\pi i}\frac{1}{E-u+i\epsilon}\begin{pmatrix}
        u^2&iu\\-iu&1
    \end{pmatrix}
    \frac{1}{u^2-\Omega^2}f(W)\\[10pt]
    &=\frac{f(W)}{2}\oint_{\mathcal{C}}\frac{du}{\pi i}\frac{1}{E-u+i\epsilon}\begin{pmatrix}
        u^2&iu\\-iu&1
    \end{pmatrix}
    \left[\frac{1}{u^2-W^2}+\frac{D}{1-D\Big(q^T\frac{W}{u^2-W^2}q\Big)}\frac{W^{1/2}}{u^2-W^2}qq^T\frac{W^{1/2}}{u^2-W^2}\right]f(W),
\end{split}\end{align}
where the contour $\mathcal{C}$ encloses the real axis counterclockwise. Then, for instance,
\begin{align}
    \begin{split}
        D\braket{\phi|\hat{B}G_-(E)\hat{B}^\dagger|\phi}&=2D\begin{pmatrix}
            0&q^T
        \end{pmatrix}\tilde{G}(E)\begin{pmatrix}
            0\\q
        \end{pmatrix}\\[5pt]
&=\oint_{\mathcal{C}}\frac{du}{\pi i}\frac{1}{E-u+i\epsilon}\times\frac{1}{1-D\Big(q^T\frac{W}{u^2-W^2}q\Big)}\\[5pt]
&=-1+\frac{2}{\pi}\int_0^{\infty} du \frac{E}{E^2+u^2}\frac{1}{1+D\Big(q^T\frac{W}{E^2+W^2}q\Big)}+\frac{2\theta(E)}{1-D\Big(q^T\frac{W}{(E+i\epsilon)^2-W^2}q\Big)}\\[5pt]
&=-1-m(E/D)+\frac{2\theta(E)}{1-D\Big(q^T\frac{W}{(E+i\epsilon)^2-W^2}q\Big)}.
    \end{split}
\end{align}
We similarly can calculate the amputated on-shell amplitude (which picks the pole after rotation of the $u$ contour $\mathcal{C}$ to the imaginary axis)
\begin{align}
    \begin{split}
        (E-\omega_k)(E-\omega_{k'})\braket{\phi|\hat{a}_kG_-(E)\hat{a}_{k'}|\phi}&=(E-\omega_k)(E-\omega_{k'})\frac{\begin{pmatrix}
            1&i
        \end{pmatrix}}{\sqrt{2}}\tilde{G}_{kk'}(E)\begin{pmatrix}
            1\\
            -i
        \end{pmatrix}/\sqrt{2}\bigg|_{\omega_{k},\omega_{k'}\to E}\\[5pt]
        &\hspace{-2.5cm}=\frac{1}{2}\begin{pmatrix}
            1&i
        \end{pmatrix} \frac{f(E)}{2}\left(\frac{2\pi i}{\pi i}\right)\begin{pmatrix}
            E^2&iE\\-iE&1
        \end{pmatrix}\frac{E^{1/2}}{2E}\frac{q_{k}^2}{1-D\Big(q^T\frac{W}{(E+i\epsilon)^2-W^2}q\Big)}\frac{E^{1/2}}{2E}f(E)\begin{pmatrix}
            1\\-i
        \end{pmatrix}\\[10pt]
        (E-\omega_k)(E-\omega_{k'})\braket{\phi|\hat{a}_kG_-(E)\hat{a}_{k'}|\phi}]\bigg|_{\omega_k,\omega_{k'}\to E} &=\frac{Dq_k^2/2}{1-D\Big(q^T\frac{W}{(E+i\epsilon)^2-W^2}q\Big)}
    \end{split}
\end{align}
Finally, we also need the amplitude of the spin-flip contributions
\begin{align}\begin{split}
    (E-\omega_k)\braket{\hat{a}_kG_-(E)\hat{B}^\dagger}&=-\sqrt{2}i(E-\omega_k)\frac{\begin{pmatrix}
        1&i
    \end{pmatrix}}{\sqrt{2}}\tilde{G}_k(E)\begin{pmatrix}
        0\\ q
    \end{pmatrix}\bigg|_{\omega_k\to E}\\[5pt]
    &=-i\begin{pmatrix}
        1&i
    \end{pmatrix}\frac{f(E)}{2}\left(\frac{2\pi i}{\pi i}\right)\begin{pmatrix}
        E^2&iE\\-iE &1
    \end{pmatrix} \begin{pmatrix}
        0\\1
    \end{pmatrix}\frac{E^{1/2}}{2E}\frac{q_k}{1-D\Big(q^T\frac{W}{(E+i\epsilon)^2-W^2}q\Big)}\\[10pt]
    (E-\omega_k)\braket{\hat{a}_kG_-(E)\hat{B}^\dagger}&=\frac{q_k}{1-D\Big(q^T\frac{W}{(E+i\epsilon)^2-W^2}q\Big)}
\end{split}\end{align}
Hence the full amputated Green's function is
\begin{align}\begin{split}
    &(E-\omega_k)(E-\omega_{k'})G_{kk'}(E)\Big|_{\omega_k,\omega_{k'}\to E}\\[10pt]&=\frac{Dq_k^2/2}{1-D\Big(q^T\frac{W}{(E+i\epsilon)^2-W^2}q\Big)}\left\{1+\frac{2D}{\Big[1-D\Big(q^T\frac{W}{(E+i\epsilon)^2-W^2}q\Big)\Big]\Big[E+Dm(E/D)\Big]-2D}\right\}
\end{split}\end{align}
The resulting $1\to 1$ $S$-matrix is given by
\begin{equation}
    S(E,E')=2\pi\delta(E-E')\frac{\Big[y+m(y)\Big]\left[1-2\alpha_D\int'\frac{x^{s+1}dx}{\big[m(x)+x\big]^2(y^2-x^2)}-\frac{i\pi\alpha_D y^s}{(y+m(y))^2}\right]-2}{\Big[y+m(y)\Big]\left[1-2\alpha_D\int'\frac{x^{s+1}dx}{\big[m(x)+x\big]^2(y^2-x^2)}+\frac{i\pi\alpha_D y^s}{(y+m(y))^2}\right]-2},
\end{equation}
which is just a phase shift and the primed integrals are principal values. We can extract the phase shift in various regimes ($\omega\ll\omega_*,\,\omega_*\ll\omega\ll D,\,D\gg \omega$). At low energies, ($y\ll 1\leftrightarrow\omega\ll D$), the $m(y)$ in denominators dominate, so
\begin{equation}
    S(E,E')\approx 2\pi\delta(E-E')\times\frac{-2\alpha_D\int'\frac{x^{s+1}dx}{\big[m(x)\big]^2(y^2-x^2)}-\frac{i\pi\alpha_D y^s}{\big[m(y)\big]^2}-\frac{2}{m(y)}}{-2\alpha_D\int'\frac{x^{s+1}dx}{\big[m(x)\big]^2(y^2-x^2)}+\frac{i\pi\alpha_D y^s}{\big[m(y)\big]^2}-\frac{2}{m(y)}},
\end{equation}
In the critical regime $\omega_*\ll\omega\ll D$ this expression is dominated by $m(y)\approx \pi \alpha_D\cot(\pi s/2)y ^s$. In the non-critical regime, $\omega\ll\omega_*$ the phase shift is dominated by $m(y)\approx m_0+2\pi \alpha_Dy^s/\sin(\pi s)$. This leads to
\begin{equation}
    S(E,E')=2\pi\delta(E-E')\times\begin{cases}
        e^{i\pi s}&\omega_*\ll\omega\ll D\\
        1+2\pi i(\omega/\omega_*)^2&\omega\ll\omega_*
    \end{cases}
\end{equation} 

\section{SCATTERING USING CFT METHODS}\label{appI}

Here we calculate the scattering matrix element in the $1\to 1$ particle channel using CFT methods. To do this, we adopt the LSZ formalism developed in~\cite{Zarand2003,Borda2007}, but focus again on the retarded propagator
\begin{equation}
    G_{\beta}=-i\theta(t)\braket{\text{gnd}|\big[\hat{a}_\beta(t),\hat{a}_{\beta}^\dagger\big]|\text{gnd}}.
\end{equation}
Using Heisenberg equations of motion, we find the following exact relations
\begin{align}
    \begin{split}
        \dot{G}_\beta&=-i\omega_\beta G_\beta-ig_\beta F_\beta-i\delta(t-t')\\
        \dot{F}_\beta&=-i\omega_\beta F_\beta-ig_\beta ZZ(t),
    \end{split}
\end{align}
where $F_\beta(t)=-i\braket{\text{gnd}|\big[\hat{s}_z(t),\hat{a}_{\beta}^\dagger\big]|\text{gnd}}$ and $ZZ(t)=-i\braket{\text{gnd}|\big[\hat{s}_z(t),\hat{s}_z\big]|\text{gnd}}$. In frequency space, we thus obtain the relation (we are omitting the $i\epsilon$ prescriptions since they will not modify the result)
\begin{equation}
    G_\beta(\Omega)=\frac{g_\beta^2}{(\Omega-\omega_\beta)^2}ZZ(\Omega)-\frac{1}{\Omega-\omega_\beta}.
\end{equation}
Thus, the relevant $T$ matrix element will be given by
\begin{equation}
    T(E)=-2\pi\delta(E-E') 2\pi\times2\alpha\omega_c^{1-s}E^s\times ZZ(E).
\end{equation}
The correlation function $ZZ(E)$ involves operators at the boundary, so at the critical point it should display power-law behaviour determined by the underlying CFT. It turns out that $\hat{s}_z$ has leading contributions from an operator with protected scaling dimension, fixed at $(1-s)/2$~\cite{Benedetti2025,Benedetti2026}. Hence, in Euclidean time, real time and frequency we have
\begin{equation}
    ZZ(\tau)=\frac{C}{\tau^{1-s}}\to ZZ(t)=-\frac{2C\sin\big[\pi(1-s)/2\big]\theta(t)}{t^{1-s}}\to ZZ(E)=-\frac{2C\sin\big[\pi(1-s)/2\big]\Gamma(s)}{E^s}e^{i\pi s/2},
\end{equation}
where $C$ is a real constant. In consequence
\begin{equation}
    S_{E,E'}=2\pi\delta(E-E')\left(1+iC{'}e^{i\pi s/2}\right),
\end{equation} 
Although the coefficient $C'>0$ is undetermined, the phase shift of the $T$ matrix is fixed by conformal invariance. If we further demand that there is no multiparticle scattering then $C'$ is fixed to $C'=2\sin(\pi s/2)$, leading to $S(E,E')=2\pi\delta(E-E') e^{i\pi s}$

\section{RELATION BETWEEN SPIN ORIENTATION AND M(0)}\label{appJ}

Here we show that the equation resulting from minimization over angles
\begin{equation}
    (\cos\theta)^2=2b\int_0^{y_c}\frac{y^{s-1}m(y)^2dy}{\big[y+m(y)\big]^2}
\end{equation}
can be recasted as $(\sin\theta)^2=m_0$. We begin from the integral equation for $m(y)$
\begin{equation}\label{eqn:appendix:mforangle}
    m(y)=\int_0^{\infty}\frac{2ydz}{y^2+z^2}\times \frac{1}{1+2bJ(z)},
\end{equation}
where we have defined
\begin{equation}
    J(z)=\int_0^{y_c}\frac{x^{s+1}dx}{(x^2+z^2)\big[x^2+m(x)\big]^2}.
\end{equation}
Plugging Eq.~(\ref{eqn:appendix:mforangle}) into the equation for the angle (only in the numerator) leads to
\begin{equation}
     (\cos\theta)^2=2b\int_0^{y_c}\frac{y^{s-1}dy}{\big[y+m(y)\big]^2}\int_0^{\infty}dz_1\int_0^{\infty} dz_2 \frac{y dz_1 ydz_2}{(y^2+z_1^2)(y^2+z_2^2)}\times\frac{1}{\big[1+2b J(z_1)\big]\big[1+2b J(z_2)]}.
\end{equation}
Rationalizing the denominators, switching the order of the integrals and doing the $y$ integral, we get
\begin{align}\begin{split}
     (\cos\theta)^2&=2b \int_0^{\infty}dz_1\int_0^{\infty} dz_2\frac{dz_1 dz_2}{(z_1^2-z_2^2)}\frac{J(z_1)-J(z_2)}{\big[1+2b J(z_1)\big]\big[1+2b J(z_2)]}\\[5pt]
     &=\int_0^{\infty}dz_1\int_0^{\infty} dz_2\frac{dz_1 dz_2}{(z_2^2-z_1^2)}\left[\frac{1}{1+2b J(z_2)}-\frac{1}{1+2bJ(z_1)}\right].
\end{split}\end{align}
The latter form evaluates to
\begin{equation}
    (\cos\theta)^2=1-\frac{1}{1+2b J(0)}.
\end{equation}
Now we note that if we $z\to zy$ in Eq.(~\ref{eqn:appendix:mforangle}) and then set $y=0$, we obtain the relation
\begin{equation}
    m(0)=m_0=\frac{1}{1+2b J(0)}=1-(\cos\theta)^2=(\sin\theta)^2,
\end{equation}
which is what we set out to demonstrate.

\end{document}